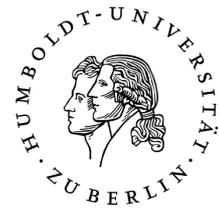

# Evaluation des Algorithmus DIBADAWN zum Detektieren von Brücken und Gelenkpunkten in 802.11 Maschennetzen

Bachelorarbeit

zur Erlangung des akademischen Grades
Bachelor of Science (B. Sc.)

| | |
|---|---|
| eingereicht von: | Robert Döring |
| geboren am: | 2.9.1984 |
| geboren in: | Bergen auf Rügen |
| Gutachter/innen: | Prof. Dr. rer. nat. Jens-Peter Redlich |
| | Prof. Dr. sc. Joachim Fischer |
| eingereicht am: ................... | verteidigt am: ................... |

Gegenstand dieser Bachelorarbeit ist die Evaluation des Algorithmus DIBADAWN zur Erkennung von Brücken und Gelenkpunkten in drahtlosen Maschennetzen. Hierfür wird der Algorithmus DIBADAWN implementiert, mittels Simulation evaluiert und die Ergebnisse mit einem Experiment verglichen.

# Inhaltsverzeichnis

# Abbildungsverzeichnis

# 1 Einleitung

Als 1999 der erste WLAN-Standard 802.11™ von der IEEE Inc. veröffentlicht wurde und erste Geräte diesen Standard unterstützen, interessierte sich kaum ein Anwender dafür. Drahtlose Netzwerke und die damit verbundene Mobilität war für die meisten Anwender uninteressant, da diese Mobilität für sie nur bedeutet hätte, dass sie ihren schweren stationären Computer samt CRT-Monitor und weiterem Zubehör von einem Platz zu einem anderen Platz innerhalb ihres drahtlosen Netzwerkes tragen können. Dabei hätten sie sich natürlich die Arbeit gespart, das Netzwerkkabel zu verlegen. Diese Arbeitsersparnis konnte allerdings nur wenige Anwender über den höheren Anschaffungspreis und die geringere Datenrate, Sicherheit und Zuverlässigkeit eines drahtlosen Netzwerkes hinwegtrösten.

Dies änderte sich um 2001, als erschwingliche Laptops auf dem Massenmarkt kamen. Mit der Verbreitung dieser Computer, welche mit dem Fokus auf Mobilität konzipiert wurden, stieg der Bedarf von Mobilität und der Bedarf von mobiler Vernetzung, welcher bis heute wachsend ist. Der WLAN-Standard 802.11™ konnte auf der einen Seite den Hunger der Anwender nach mobiler Vernetzung sättigen und wurde auf der anderen Seite stetig durch die Ansprüche der Anwender erweitert. So wurden, unter anderem, Aspekte, wie Sicherheit, Geschwindigkeit und Zuverlässigkeit mit Teilspezifikationen, wie 802.11-ac, ergänzt.

Weiter wurde mit der Teilspezifikation 802.11s™ die Unterstützung für vermaschte drahtlose Netzwerke in der OSI-Sicherungsschicht des Standards hinzugefügt, was den Bedarf an effizienten Maschennetzen widerspiegelte [Soc12]. Maschennetze sind Netzwerke ohne einen zentralen administrierenden Teilnehmer, welcher die Funktionen im Netzwerk steuert. Jeder Teilnehmer im Maschennetz ist mit mindestens einem weiteren Teilnehmer verbunden und somit im Maschennetz integriert. Ein Teilnehmer eines Maschennetzes kann andere Teilnehmer X erreichen, indem er über eine Liste von weiteren Teilnehmern mit diesem Teilnehmer X verbunden ist.

Dies impliziert dass alle Teilnehmer miteinander verbunden sein können und es somit möglich wäre, dass alle Teilnehmer im Maschennetz direkt miteinander kommunizieren könnten. Eine direkte Kommunikation hört sich auf dem ersten Blick vielversprechend an. Es gäbe keine Latenz durch das Weiterleiten und ein komplexes Routing wäre auch nicht mehr erforderlich. Dennoch ist das Gegenteil für drahtlose Maschennetze der Fall, da sich alle Teilnehmer in einem drahtlosen Netzwerk ein gemeinsame Übertragungsmedium teilen. Ist die Teilnehmeranzahl und das Kommunikationsaufkommen genügend groß, so wird es allen Teilnehmern des drahtlosen Maschennetzes schier unmöglich überhaupt etwas zu übertragen. Eine optimale Anordnung der Teilnehmer zu erhalten ist keine triviale Aufgabe. Ein Teilnehmer muss ausreichend benachbarte Teilnehmer besitzen, um weiterhin mit dem Maschennetz verbunden zu sein, falls eine Menge von Teilnehmern aus diesem Maschennetz ausscheidet. An dieser Stelle sind noch eine Vielzahl weiterer Faktoren zu berücksichtigen, wie etwa der maximale Durchsatz im gesamten Netz und der einzelnen Teilnehmer, die Fairness beim Mediumzugriff, die zu erzielende Mobilität der Teilnehmer und der Energieaufwand für Nachrichtenübertragungen.

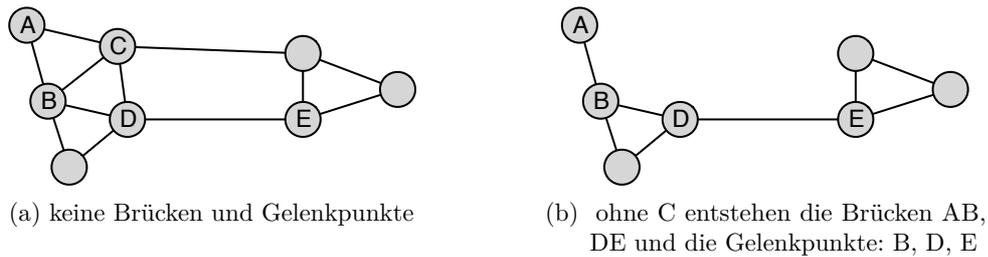

(a) keine Brücken und Gelenkpunkte  (b) ohne C entstehen die Brücken AB, DE und die Gelenkpunkte: B, D, E

Abbildung 1.1: Entstehung von Brücken und Gelenkpunkten

Es lässt sich vorstellen, dass eine Anordnung als theoretisches Optimum für ein konkretes Szenario ermittelt wird, sei es durch eine Berechnung oder mittels einer Simulation. Dennoch lässt es sich ebenso denken, dass es in der Realität andere Faktoren dieser optimalen Anordnung der Netzwerkteilnehmer entgegen steht. Will man beispielsweise bestehende Windkraftanlagen in einem Offshore-Windpark mit einem drahtlosen Maschennetzwerk miteinander vernetzen, so hat man kaum Einfluss auf die Anordnung der einzelnen Teilnehmer, da die Position der Windkraftanlagen zuvor bestimmt wurde.

Gerade in den Fällen, in denen keine optimale Anordnung der Netzwerkteilnehmer besteht, ist das Erkennen von Gelenkpunkten, also den Teilnehmern, ohne welche mindestens zwei weitere Teilnehmer nicht mehr miteinander kommunizieren könnten, von Bedeutung. Das gleiche gilt für Brücken, also Verbindungen zwischen Netzwerkteilnehmern, ohne welche mindestens zwei Teilnehmer nicht mehr miteinander kommunizieren könnten. Scheidet exemplarisch der Teilnehmer C aus dem Maschennetz aus Abbildung 1.1a aus, so ist es möglich, dass weiterhin alle anderen Teilnehmer miteinander kommunizieren können, da es die Brücken AB, DE und die Gelenkpunkte B,D und E noch gibt. Es kann also zu spontanen Ausfällen von Netzwerkteilnehmern oder Verbindungen zwischen diesen kommen. In solchen Fällen könnte die Brücken und Gelenkpunkterkennung dafür genutzt werden, um auf die entstehenden Brücken und Gelenkpunkte als potentielle Schwachstellen im Maschennetzwerk hinzuweisen und somit eine Spaltung des Maschennetzes im Vorfeld zu verhindern.

Ist es nicht möglich die Existenz von Brücken oder Gelenkpunkten zu verhindern, so kann bereits das Wissen über die Existenz einen Vorteil bringen. Möchte ein Netzwerkteilnehmer eine Nachricht zu einem anderen Netzwerkteilnehmer in einem drahtlosen Maschennetzwerk versenden, so gibt es dafür mehrere mögliche Pfade, wenn zwischen diesen Netzwerkteilnehmern keine Brücke existiert. Besteht jedoch ein Brücke zwischen ihnen, so muss die Nachricht über eben genau diese Brücke übertragen werden und sollte dabei besser nicht verloren gehen. Hierfür könnte beispielsweise der an eine Brücke angrenzende Absender seine Sendeleistung erhöhen oder eine robustere Übertragungsmöglichkeit wählen, um somit die Übertragungswahrscheinlichkeit gezielt zu erhöhen.

## 1.1 Schwierigkeit bei der Erkennung der Topologie

Die Kommunikation in einem Wireless Multi-Hop Netzwerk muss mit verschieden Widrigkeiten umgehen. Im Gegensatz zu Teilnehmern in kabelgebundenen Netzwerken teilen sich alle Teilnehmer in einem drahtlosen Netzwerk ein gemeinsames Übertragungsmedium, und dieses besitzt eine beschränke Bandbreite. Somit erhöht sich die Wahrscheinlichkeit einer Kollision bei einer Nachrichtenübertragung mit der Teilnehmeranzahl und dem Kommunikationsaufkommen im Netzwerk. Ebenso können Umwelteinflüsse, wie das Wetter, die Vegetation, die Entfernung zwischen den Teilnehmern oder netzwerkfremde technische Geräte die Signalausbreitung nachteilig beeinträchtigen und somit einen Nachrichtenverlust bewirken. Aus diesen Gründen ist der Erfolg einer Kommunikation in einem drahtlosen Netzwerk nicht vorhersagbar.

Im Standard 802.11-12™ der IEEE Inc. ist das Erkennen einer fehlerhaften Unicast-Übertragung durch das Ausbleiben einer Bestätigungsnachricht [Soc12, S. 824] beschrieben. Beim Versenden von Broadcast-Nachrichten ist dem Absender nicht bekannt, welche Teilnehmer diese Nachricht empfangen. Selbst wenn der Absender Informationen über seine benachbarten Teilnehmer besitzt, so kann er nicht ausschließen, dass genau in dem Moment des Versendens ein neuer Teilnehmer im Netzwerk hinzukommt oder ein bekannter aus seiner Nachbarschaft entfernt wird. Daher ist das Warten auf genau eine Bestätigungsnachricht und das Warten auf alle Bestätigungsnachrichten der bekannten Teilnehmer in der Nachbarschaft nicht zielführend, um den Erfolg beziehungsweise Misserfolg einer Broadcast-Übertragung nachzuweisen. Somit kann ein Absender einer Broadcast-Nachricht nicht erkennen, ob seine versendete Nachricht fehlerhaft übertragen wurde, was beispielsweise aufgrund von Kollisionen vorkommen kann.

Da Kollisionen in einem drahtlosen Netzwerk auftreten können aber vom Absender nicht zuverlässig erkannt werden, gilt eine Übertragung in einem Wireless Multi-Hop Netzwerk als unzuverlässig. Eine unzuverlässige Nachrichtenübertragung ist eine denkbar schlechte Ausgangssituation, um Verbindungen zwischen Netzwerkteilnehmern in einem drahtlosen Netzwerk genau zu bestimmen.

Von einem Algorithmus für die Erkennung von Brücken und Gelenkpunkten wird demnach erwartet, dass er die schwankenden Informationen über die Verbindungen zwischen den Teilnehmern im Netzwerk so stabilisiert, dass die Erkenntnisse, welche er auf der Grundlage dieser Informationen erhält, möglichst zutreffend sein werden.

## 1.2 Netze als Graphen

Eines der vermutlich am häufigsten in der Informatik untersuchten Strukturen sind Graphen. Ein Graph, also eine Menge von Knoten mit einer Menge von Verbindungen zwischen diesen Knoten, kann ebenfalls als eine abstrakte Darstellung eines komplexen Sachverhalts verstanden werden. Sodass ein komplexer Sachverhalt auf ein Graphenproblem zurückgeführt werden kann.

So auch in diesem Fall. Der komplexe zu betrachtende Sachverhalt ist das drahtlose Maschennetzwerk, welches durch mehrere tausend Seiten lange Standards spezifiziert ist. Dieser Sachverhalt wird auf ein Graphenproblem zurückgeführt, indem alle Teilneh-mer eines drahtlosen Maschennetzes eine Menge von Knoten und alle Verbindungen zwischen diesen Teilnehmern eine Menge an Verbindungen bilden. In diesem Graphen können Brücken und Gelenkpunkte gesucht werden, welche durch den zu evaluierenden Algorithmus DIBADAWN ebenfalls erkannt werden sollen ohne über die Kenntnis der Topologie zu verfügen.

Dies ist der Grund dafür, dass der Begriff Netzwerkteilnehmer als Synonym für den Begriff Knoten und der Begriff Netzwerk als Synonym für den Begriff Graph in dieser Arbeit verwendet wird.

### 1.3 Aufbau dieser Arbeit

Diese Arbeit gliedert sich grob in zwei Teile. Im ersten Teil wird die Problematik der Brücken- und Gelenkpunkterkennung allgemein vorgestellt, bisherige Ansätze betrachtet und der hier evaluierte Algorithmus DIBADAWN vorgestellt. Im zweiten Teil wird auf die Implementierung, die Vorgehensweise bei der Evaluation und auf die Ergebnisse der Evaluation eingegangen.

Im Abschnitt 2 werden allgemeine Definitionen und Abkürzungen aufgelistet, welche in der gesamten Arbeit verwendet werden. Im Abschnitt 3 werden andere Arbeiten vorgestellt, welche sich ebenfalls mit dem Thema der Brücken- oder Gelenkpunkterkennung befassen oder relevant sind für das Verständnis des Algorithmus DIBADAWN. Im Abschnitt 4 ist der hier behandelte Algorithmus DIBADAWN beschrieben. Im Abschnitt 5 ist die Implementation beschrieben, welche in dieser Arbeit verwendet wurde, um den Algorithmus DIBADAWN zu evaluieren. Im Abschnitt 6 wird die Methodik beschrieben, mit welcher die Evaluation durchgeführt wird. Im Abschnitt 7 sind die evaluierten Versuchsaufbauten und die Ergebnisse der Evaluationen aufgeführt. Abschließend wird im Abschnitt 8 eine Zusammenfassung der vorliegenden Arbeit gegeben und mögliche Verbesserungen werden vorgestellt.

## 2 Definitionen und Abkürzungen

### 2.1 Definitionen

**Definition 2.2** (Verbindungstopologie)**.** In einem Netzwerk bezeichnet man die Anordnung der Netzwerkteilnehmer und deren Verbundenheit als Verbindungstopologie. [Sch05, S. 187]

**Definition 2.3** (Komponente)**.** Sei G ein ungerichteter Graph derart, dass zwischen zwei beliebigen Knoten aus G ein Pfad existiert, so nennt man diesen zusammenhängend. Sei ein Teilgraph $G_T$ maximal zusammenhängend, also zwischen allen Knoten A und B aus $G_T$ existiert ein Pfad, so nennt man $G_T$ eine Komponente von G.

**Definition 2.4** (Brücke)**.** Eine Kante in einem ungerichteten Graphen G ohne welche die Anzahl der Komponenten in G steigt, nennt man Brücke.

**Definition 2.5** (Gelenkpunkt)**.** Einen Knoten in einem ungerichteten Graphen G, ohne welchen die Anzahl der Komponenten in G steigt, nennt man Gelenkpunkt.

**Definition 2.6** (unidirektionaler Link)**.** In einem gerichteten Graphen besteht ein bidirektionaler Link zwischen zwei Knoten A und B genau dann, wenn eine gerichtete Kante von A zu B existiert. In einem Netzwerk besteht ein bidirektionaler Link genau dann zwischen zwei Netzwerkteilnehmern A und B, wenn Daten von A zu B übertragen werden können.

**Definition 2.7** (bidirektionaler Link)**.** In einem ungerichteten Graphen besitzen zwei Knoten genau dann einen Link, wenn eine ungerichtete Kante zwischen diesen Knoten existiert. In einem Netzwerk besitzen zwei Netzwerkteilnehmer genau dann einen Link, wenn Daten zwischen diesen Netzwerkteilnehmern in beide Richtungen übertragen werden können.

**Definition 2.8** (Link)**.** In einem Graphen besitzen zwei Knoten einen Link, wenn diese Knoten einen bidirektionalen- oder unidirektionalen Link besitzen. In einem Netzwerk besitzen zwei Netzwerkteilnehmer einen Link, wenn diese einen bidirektionalen- oder unidirektionalen Link besitzen.

**Definition 2.9** (ETX-Link)**.** Sei $E_S$ ein ETX-Schwellwert. In einem Graphen besitzen zwei Knoten einen ETX-Link, wenn diese Knoten einen 1-Hop Link besitzen und dieser Link eine ETX-Metrik $E_L$ besitzt, sodass gilt: $E_L \leq E_S$. In einem Netzwerk besitzen zwei Netzwerkteilnehmer einen ETX-Link, wenn diese einen 1-Hop Link mit einer ETX-Metrik $E_L$ besitzen, sodass gilt: $E_L \leq E_S$.

**Definition 2.10** (ETX-Topologie)**.** Sei $E_S$ ein ETX-Schwellwert. Eine ETX-Topologie ist eine Topologie in der für alle Links mit einer ETX-Metrik $E_L$ gilt, dass $E_L \leq E_S$.

**Definition 2.11** (Spannbaum)**.** Sei G ein Graph mit genau einer Komponente. Ein zusammenhängender, zyklenfreier Teilbaum von G heißt Spannbaum.

**Definition 2.12** (Kreis)**.** Sei $G := (V, E)$ ein ungerichteter Graph. Ein Pfad $P \in V^n$ : $(v_0, v_1, .., v_n)$ mit

$$
\begin{aligned}
&(1) \quad n \geq 2 \\
&(2) \quad v_0 = v_n \\
&(3) \quad v_i \neq v_j \quad \text{für } i, j \in \{1, .., n\} : i \neq j
\end{aligned}
$$

nennt man Kreis.

**Definition 2.13** (Kreisrelation)**.** Sei $G := (V, E)$ ein ungerichteter Graph. Zwei Kanten $A, B \in E : A \neq B$ befinden sich in einer Kreisrelation, wenn ein Kreis $C := (V', E')$ mit $V' \subseteq V, E' \subseteq E$ existiert, sodass gilt: $A, B \in E'$. [Mil09, S. 80]

## 2.14 Abkürzungen

| | |
|---:|---|
| CRT | Cathode Ray Tube |
| DIBADAWN | Distributed Bridge and Articulation Point Detection Algorithm for Wireless Networks |
| ETX | Expected Transmission Count |
| IEEE | Institute of Electrical and Electronics Engineers |
| MAC | Medium Access Control |
| OSI-Modell | Open Systems Interconnection Model |
| WLAN | Wireless Local Area Network |

# 3 Verwandte Arbeiten

Neben dem hier behandelten Algorithmus DIBADAWN existieren weitere Algorithmen zur Erkennung von Brücken und Gelenkpunkten. In diesem Abschnitt sollen exemplarisch zwei Vertreter vorgestellt werden. Eine weitere verwandte Arbeit ist der Echo-Algorithmus [Cha82] und natürlich die Arbeit des Autors von dem hier evaluierten Algorithmus DIBADAWN [Mil09].

## 3.1 Robert Tarjans Algorithmus

Einer der wohl bekanntesten Algorithmen für die Suche von Gelenkpunkten in Graphen ist der Algorithmus von Robert Tarjan aus dem Jahr 1972 [Tar72], auch bekannt als Tarjans Algorithmus, für die Suche von Gelenkpunkten. Die Grundidee bei diesem Algorithmus ist das Suchen von Kreisen beim Erstellen eines Spannbaumes unter der Verwendung der Tiefensuche, um Gelenkpunkte in einem Graphen zu finden.

Bei diesem Algorithmus wird von einem Graphenproblem ausgegangen. Das bedeutet, es sei ein Graph gegeben und Gelenkpunkte in diesem Graphen werden gesucht. Bei dieser Suche wird auf das Wissen über diesen Graphen zurückgegriffen. Übertragen auf ein drahtlosen Maschennetzwerk bedeutet dies, dass alle Knoten über ein globales Wissen über das Netzwerk verfügen, was sehr aufwendig zu realisieren sein dürfte. Dennoch soll dieser Algorithmus hier beschrieben werden, da er der Urahne aller Algorithmen für die Gelenkpunktsuche ist.

Ein zentrales Element bei Tarjans Algorithmus ist ein gemeinsamer Zähler. Er wird in der Zeile 1 so initialisiert, dass er in der Zeile 2 als erstes die Zahl Eins liefert und nachfolgend die Zahlen Zwei, Drei und so weiter. Gegeben sei ein Graph $G := (V, E)$. Der Algorithmus wird für alle Knoten $v \in V$ ausgeführt, welche bisher nicht nummeriert wurden, wie in Zeile 7 und 8 dargestellt.

Im ersten Schritt wird dem ausgewählten Knoten $v$ mit dem beschrieben Zähler eine eindeutige Zahl zugewiesen, wie in Zeile 2 zu sehen. Jedem nummerierten Knoten wird eine minimale Zahl zugeordnet, diese Zahl entspricht der kleinsten Nummer im erkannten Spannbaum. Die ihm bekannten Zahlen sind die Zahlen all seiner Kindknoten und seine eigene Zahl. Anfangs besitzt ein Knoten $v$ keine Kindknoten und daher wird

**Algorithmus 1 :** Tiefensuche aus Tarjans Algorithmus für die Gelenkpunktsuche

**Data** : Graph $G := (V, E)$
1 **Data** : counter := newSequentialCounter().initNextNumber(1)
**Data** : stack := newEmptyStack()

**input** : v := ein Knoten
**input** : parent := Elternknoten im Suchbaum
**Function** *BICONNECT(v, parent)*
2     v.setNumber(counter.next());
3     v.setMin(v.getNumber());
    **for** $w \in v.listOfAdjacent()$ **do**
       **if** *w.isNotNumbered()* **then**
          stack.add(v, w);
4           BICONNECT(w,v);
          v.setMin(min(v.getMin(), w.getMin()));
5           **if** *w.getMin() $\geq$ v.getNumber()* **then**
             component = newComponent();
             **while** $(u_1, u_2) := stack.top()$ *mit* $u_1.getNumber() \geq w.getNumber()$ **do**
                stack.remove($u_1, u_2$);
                component.add($u_1, u_2$);
             stack.remove($v, w$);
             component.add($v, w$);
6        **else if** *w.getNumber() < v.getNumber() AND $w \neq$ parent* **then**
          stack.add(v, w);
          v.setMin(min(v.getMin(), w.getMin()));

/* Initialer Aufruf                                                                                     */
**for** $v \in V$ **do**
7     **if** *v.isNotNumbered()* **then**
8        BICONNECT(v,NULL);

die minimale Zahl von $v$ auf die Zahl von $v$ selbst gesetzt, Zeile 3. Als nächstes wird eine Liste der Nachbarn von $v$ betrachtet. Hierfür wird ein Nachbar $w$ von $v$ gewählt.

Wurde der Nachbar $w$ bereits vor $v$ mit einer kleineren Zahl nummeriert und ist $w$ nicht der Vorgänger, also parent, von $v$ im entstehenden Spannbaum, so wird die Kante $(v, w)$ auf den Stack gelegt und die minimale Zahl von $v$ auf die minimale Zahl von $w$ aktualisiert. Auf diese Weise wird ein Kreis im Graphen gefunden und die letzten Kanten auf dem Stack gehören zu einer gemeinsamen Komponente.

Wurde dieser Nachbar $w$ noch nicht nummeriert, so wird die Kante $(v, w)$ auf einem gemeinsamen Stack abgelegt und es werden alle Schritte ab dem ersten Schritt für den Knoten $w$ mit dem Knoten $v$ als parent wiederholt, bevor der Algorithmus an dieser Stelle fortgesetzt wird, wie in den Zeilen 4 dargestellt. Somit entsteht auf dem Stack nach und nach ein Spannbaum. Nun kennt $v$ den Knoten $w$ als Kindknoten und die minimale Zahl von $v$ wird auf den minimalen Wert von $w$ gesetzt, wenn dieser Wert kleiner als der bisherige minimale Wert von $v$ ist. Ist der minimale Wert von $w$ größer oder gleich der Zahl von $v$, so ist $v$ der Knoten zu welchem zuvor in einer anderen Rekursion in der Zeile 6 ein Kreis erkannt wurde. Das bedeutet, dass alle Kanten auf dem Stack ab der Kante, welche $v$ als Startknoten enthält, zu einer gemeinsamen Komponente gehören. All diese Kanten werden im Anweisungsblock der Bedingung in Zeile 5 in einer gemeinsame Komponente verschoben und somit wieder vom Stack entfernt.

Im eigentlichen Algorithmus von Tarjan werden Gelenkpunkte zwar nicht explizit als solche benannt, aber jeder Knoten in einer Komponente mit der kleinsten Zahl ist entweder ein Gelenkpunkt oder die Wurzel im entstandenen Spannbaum [Tar72, S. 151].

## 3.2 Verteilter Algorithmus

In diesem Abschnitt wird ein verteilter Algorithmus von Pranay Chaudhuri vorgestellt, welcher Gelenkpunkte in einem Netzwerk erkennt. Der Algorithmus wurde 1998 veröffentlicht und findet Gelenkpunkte in $\mathcal{O}(n)$ Zeiteinheiten unter Verwendung von $\mathcal{O}(n)$ Nachrichten mit $n$ Knoten im Netzwerk [Cha98].

Bei diesem Algorithmus ist allerdings zu beachten, dass einige Annahmen getroffen werden, welche die Nutzung des Algorithmus für drahtlose Netzwerke ausschließt. Zum Einen werden hohe Anforderungen an die Links gestellt. Es wird davon ausgegangen, dass die Links im Netzwerk bidirektional sind. Dass heißt, wenn ein Knoten A zu einem Knoten B Daten übertragen kann, dann kann auch B an A Daten übertragen. Laut einer experimentellen Untersuchung kommen solche Links in einem drahtlosen Netzwerk zwar vor, sind aber sehr viel seltener als unidirektionale Links, bei welchen eine Datenübertragung nur in genau eine Richtung möglich ist [LKN12]. Weiter soll es zwischen den Links keine Interferenzen geben, was ebenfalls nicht in drahtlosen Netzwerken gewährleistet ist. Es wird angenommen, das ein Link aus zwei Kanälen besteht, wobei ein Kanal jeweils eine Übertragungsrichtung zwischen zwei Knoten abdeckt. Die Zuverlässigkeit der Links und Kanäle soll ausreichen, um Übertragungsfehler auszuschließen. Zum Anderen wird angenommen, dass ein Knoten die Menge seiner Nachbarn kennt. Trotz dieser Annahmen ist dieser Algorithmus hier beschrieben, da er ein verteilter Algorithmus ist.

Der Algorithmus wird hier nicht im Detail beschrieben, dennoch soll die Arbeitsweise grob umrissen werden. Da es sich um einen verteilten Algorithmus handelt, ist globales Wissen, auf welches alle Knoten zugreifen können, nicht vorhanden und ein Wissensaustausch wird über das Versenden von Nachrichten ermöglicht. Der proaktive Algorithmus wird von einem beliebigen Knoten im Netzwerk initiiert und dieser Knoten nimmt während der Ausführung eine Sonderstellung ein. Nach der Ausführung weiß jeder Knoten, ob er selbst ein Gelenkpunkt ist oder nicht. Eine Verbreitung dieses Wissens ist nicht vorgesehen. Die Arbeitsweise des Algorithmus lässt sich in zwei Phasen aufteilen. In der ersten Phase wird ein Spannbaum erzeugt und in der zweiten Phase werden mit diesem Spannbaum Gelenkpunkte erkannt.

Der Spannbaum wird mit einer Tiefensuche gefunden. Hierbei wird die Tiefensuche von einem beliebigen Knoten $r$ im Netzwerk initiiert und dieser Knoten wird die Wurzel eines Spannbaumes. Im Rahmen einer Tiefensuche werden SEARCH-Nachrichten versendet, welche eine Menge der bisher besuchten Knoten beinhalten. Empfängt ein Knoten $v$ mit $v \neq r$ erstmalig eine SEARCH-Nachricht, so vermerkt er sich den Absender als parent, fügt sich selbst zur Menge der besuchten Knoten in der SEARCH-Nachricht hinzu und versendet die SEARCH-Nachricht an einen ihm bekannten Nachbarknoten, welcher entsprechend der besuchten Knoten aus der SEARCH-Nachricht noch nicht besucht wurde, und vermerkt sich diesen Nachbarknoten in einer Menge von Kinderknoten. Existiert kein Nachbarknoten, welcher noch nicht besucht wurde, so vermerkt sich der Knoten alle Links zu Nachbarknoten, welche nicht in seiner Menge der Kinderknoten enthalten sind in seiner Menge von Querverbindungen und sendet die SEARCH-Nachricht mit der aktuellen Menge der besuchten Knoten an seinen parent zurück. Diese Tiefensuche wird schrittweise fortgesetz, bis $r$ eine SEARCH-Nachricht erhält und es keinen Nachbarknoten von $r$ gibt, welcher noch nicht besucht wurde. In diesem Fall versendet $r$ eine TERMINATE-Nachricht an jeden Knoten aus der Menge seiner Kinderknoten und diese leiten wiederum die TERMINATE-Nachricht an all ihre Kinder weiter und somit terminiert die Tiefensuche in jedem Knoten und der Spannbaum gilt als gefunden.

In der zweiten Phase des Algorithmus werden alle Gelenkpunkte im Netzwerk erkannt. Der Wurzelknoten $r$ kann sofort beurteilen, ob er ein Gelenkpunkt ist, wenn er mehr als zwei Nachbarknoten hat, welche sich in seiner Menge der Kinderknoten befinden, so ist er ein Gelenkpunkt. Für alle anderen Knoten gestaltet sich diese Entscheidung etwas anders. Wenn ein Knoten $v$ mit $v \neq r$ in die zweite Phase übergeht, so schaut er zuerst, ob die Anzahl seiner Menge der Kinderknoten gleich Null ist. Wenn dies zutreffend ist, so versendet $v$ eine NONTREE-Nachricht an seinen parent. Mit einer NONTREE-Nachricht wird eine Menge von Querverbindungen versendet. Anhand der empfangenen Querverbindungen entscheidet der Empfängerknoten, ob er ein Gelenkpunkt ist. Empfängt ein Knoten eine NONTREE-Nachricht mit einer leeren Menge an Querverbindungen, so ist der Empfängerknoten ein Gelenkpunkt. Empfängt ein Knoten $v$ eine NONTREE-Nachricht mit einer Menge an Querverbindungen, sodass $v$ ein Knoten einer dieser Querverbindungen ist, so ist $v$ ein Gelenkpunkt. Hat ein Knoten $v$ eine NONTREE-Nachricht von jedem Knoten aus der Menge der Kinderknoten von $v$ erhalten, so sendet $v$ ebenfalls eine NONTREE-Nachricht an seinen parent. Für jeden Knoten, einschließlich dem Wurzelknoten, gilt: Wenn ein Knoten von jedem Knoten aus seiner Menge der Kinderknoten eine NONTREE-Nachricht erhalten hat und gegebenenfalls eine NONTREE-Nachricht in seinen parent versendet hat, so terminiert er. Das heißt, ausgehend von den Blättern des Spannbaumes werden Nachrichten in Richtung Wurzel des Spannbaumes versendet und jeder Empfängerknoten leitet diese Nachricht weiter, sobald er von all seinen Kindknoten eine Nachricht erhalten hat und terminiert seine Ausführung. Anhand der empfangenen Nachrichten entscheidet der Empfängerknoten, ob er ein Gelenkpunkt ist oder nicht. Hat der Wurzelknoten des Spannbaumes eine Nachricht von all seinen Kindernoten erhalten, so terminiert auch er seine Ausführung und der gesamte Algorithmus ist abgeschlossen und alle Gelenkpunkte sind gefunden.

# 4 Der Algorithmus DIBADAWN

Der Algorithmus **D**istributed **B**ridge and **A**rticulation Point **D**etection **A**lgorithm for **W**ireless **N**etworks, kurz DIBADAWN, wurde an der Humboldt Universität zu Berlin von Dr. Bratislav Milic im Rahmen seiner Dissertation [Mil09] entwickelt. Wie der Name vermuten lässt, dient er zur verteilten Erkennung von Brücken und Gelenkpunkten in einem drahtlosen Maschennetzwerk. DIBADAWN ist angelehnt an den Echo-Algorithmus [Cha82] und Tarjan's Algorithmus [Tar72].

Der Algorithmus selbst ist in einen Analyseteil und einen Ausführungsteil gegliedert. Im Ausführungsteil wird ausgehend von einem Knoten im Netzwerk ein Algorithmus angewendet, um Brücken und Gelenkpunkte zu erkennen. Der Analyseteil dient der Stabilisierung der Ergebnisse, da es aufgrund der erschwerten Übertragung in einem drahtlosen Netzwerk zu fehlerhaften Resultaten im Ausführungsteil kommen kann.

Wie in der Einleitung beschrieben können Netzwerke als Graphen aufgefasst werden. Der Algorithmus DIBADAWN arbeitet auf ungerichteten Graphen mit dem Ziel, alle Brücken und Gelenkpunkte in diesem Graphen zu finden.

## 4.1 Was leistet DIBADAWN?

Das Übertragungsmedium wird in einem drahtlosen Netzwerk von allen Kommunikationsteilnehmern gemeinsam genutzt und das Risiko für eine Nachrichtenkollision steigt mit dem Kommunikationsaufkommen. Daher werden für eine DIBADAWN-Ausführung in einem Graphen mit $n$ Knoten $2n - 1$ Nachrichten versendet.

DIBADAWN ist ein verteilter Algorithmus. In der Graphentheorie ist ein verteilter Algorithmus ein Menge von Turing-Maschinen im Sinne der theoretischen Informatik, wobei jede Turing-Maschine zusätzlich über eine Möglichkeit zum Senden- und Empfangen von Daten verfügt. Jeweils eine Turing-Maschine wird einem Knoten im Graphen zugeordnet [Ray13]. Im Fall von DIBADAWN bedeutet dies zum Einen, dass die Ergebnisse des Algorithmus nichtsequenziell von den Knoten eines Graphen berechnet werden und diese Knoten sich mittels Nachrichten untereinander austauschen. Zum Anderen bedeutet dies auch, dass nicht alle Knoten im Graphen über Brücken und Gelenkpunkte informiert werden. Es werden lediglich die angrenzenden Knoten über das Vorhandensein von Brücken und Gelenkpunkten informiert.

Bei der Übertragung von Nachrichten in einem drahtlosen Netzwerk kann es zu fehlerhaften Übertragungen kommen. Daher ist eine zuverlässige Link-Erkennung zwischen zwei Knoten nicht möglich [Mil09, S.45ff]. Insbesondere zwei Knoten mit einer Entfernung, welche ungefähr der maximalen Sendereichweite eines Knoten entspricht, werden Pakete teils erfolgreich und teils erfolglos übertragen. Abhängig vom Erfolg der Übertragungen wird eine unterschiedliche Verbindungstopologie erkannt (Definition Link 2.8) und somit werden auch unterschiedliche Brücken und Gelenkpunkte erkannt. Daher verwendet DIBADAWN einen statistischen Ansatz, bei welchem die vergangenen DIBADAWN-Ausführungen für die Erkennung von Brücken und Gelenkpunkten herangezogen werden.

### 4.2 Was leistet DIBADAWN nicht?

Die Vorteile von DIBADAWN werden nicht ohne Einschränkungen erhalten. Die Minimierung der zu versendenden Nachrichten bei der DIBADAWN-Ausführung hat den Preis, dass nicht alle Netzwerkteilnehmer über vorhandene Brücken und Gelenkpunkte im Netzwerk informiert werden. Es werden lediglich die angrenzenden Netzwerkteilnehmer über deren Existenz informiert. Das heißt, der DIBADAWN-Algorithmus stellt statt einem globalem Wissen ein lokales Wissen über die Existenz von Brücken und Gelenkpunkten im Netzwerk bereit. Dies könnte geändert werden, indem eine zusätzliche Komponente die Erkenntnisse einer DIBADAWN-Ausführung global im Netzwerk bereitstellt.

Die Verbindungstopologie von drahtlosen Netzwerken kann sich im Verlauf der Zeit ändern, wodurch sich ebenfalls die Existenz von Brücken und Gelenkpunkten im Netzwerk ändern kann. Eine zeitnahe Reaktion auf derartige Änderungen können wünschenswert sein. Beispielsweise könnten zwei stark verbundene Netzwerke über genau zwei Netzwerkknoten A und B miteinander verbunden sein, sodass A und B nicht miteinander verbunden sind. Würde der Netzwerkknoten A ausfallen, wäre der Netzwerkknoten B ein Gelenkpunkt. Auf derartige Veränderungen kann DIBADAWN nicht ereignisgesteuert reagieren. Der DIBADAWN-Algorithmus muss periodisch ausgeführt werden, um derartige Veränderungen zu erkennen. Das bedeutet, der DIBADAWN-Algorithmus muss umso öfter ausgeführt werden, desto schneller eine Änderung im Netzwerk erkannt werden soll. Dies kann eine hohe Belastung für ein Netzwerk darstellen, allerdings ist es denkbar, einen Teil der aufkommenden Nachrichten gebündelt mit Nachrichten anderer Protokolle zu versenden, sodass die Belastung im Netzwerk insgesamt verringert wird.

Der Algorithmus DIBADAWN geht bei der Suche von Brücken und Gelenkpunkten von einem ungerichteten Graphen aus. Übertragen auf drahtlose Maschennetze bedeutet dies, dass nur bidirektionale Verbindungen im Maschennetzwerk betrachtet werden.

Wenn ein Knoten A Daten zu einem Knoten B senden kann, dann soll auch B Daten an A übertragen können. Andernfalls ist die Verbindung zwischen A und B nicht unidirektional und es liegt nach Definition 2.8 kein Link vor. Nur wurde experimental nachgewiesen, dass unidirektionale Links, also Links mit der Möglichkeit Daten nur in genau eine Richtung zu übertragen, in drahtlosen Netzwerken 8,69 mal häufiger vorkommen als bidirektionale Links [LKN12].

## 4.3 Funktionsweise einer Ausführung

In diesem Abschnitt wird die Funktionsweise einer DIBADAWN-Ausführung beschrieben. Hierbei wird vorerst von einer fehlerfreien Nachrichtenübertragung ausgegangen, um die Funktionsweise möglichst anschaulich vermitteln zu können. Die auftretenden Probleme einer nicht-fehlerfreien Nachrichtenübertragung und bestehende Lösungsansätze sind im Abschnitt 4.6 dargelegt.

Umgangssprachlich ausgedrückt kann die im DIBADAWN-Algorithmus implementierte Suche nach Brücken und Gelenkpunkten in einem Netzwerk als ein Zyklus verstanden werden. Einem Zyklus, in dem sich die Suche ausgehend von einem Netzwerkknoten zirkulär ausbreitet und anschließend koordiniert von Außen nach Innen wieder zusammenzieht.

Ebenso wie der Echo-Algorithmus ist DIBADAWN in eine Forward-Phase und eine Backward-Phase aufgeteilt. In der Forward-Phase breitet sich die Suche im Netzwerk aus und dabei wird ein Spannbaum des Netzwerks schrittweise aufgebaut und Querverbindungen zwischen Knoten innerhalb dieses Spannbaumes ermittelt. In der Backward-Phase findet eine Auswertung dieses ermittelten Graphen statt. Dabei wird ausgehend von den Blättern des Spannbaumes zeitlich koordiniert die Suche schrittweise beendet und die gewonnenen Ergebnisse lokal vermerkt.

## 4.4 Forward-Phase

Der DIBADAWN-Algorithmus wird von einem Knoten im Graphen initiiert, dem Initiator. Er befindet sich anfänglich in der Forward-Phase und erstellt einen einzigartigen Explorer. Dieser Explorer kann als Kennung der aktuellen DIBADAWN-Ausführung verstanden werden. Der Initiator versendet diesen Explorer mit einer Explorer-Nachricht $msg$ an all seine benachbarten Knoten per Broadcast. Beim Empfang einer Explorer-Nachricht $msg_{in}$ sind folgende Fälle zu unterscheiden:

a Der empfangene Explorer ist dem Empfänger unbekannt.
   In diesem Fall merkt sich der Empfänger den Absender, $msg_{in}.parent$, der Explorer-Nachricht lokal als parent und sendet den empfangenen Explorer mit der Angabe des Absender als ursprünglichen Absender treeParent mit einer Explorer-Nachricht an all seine Nachbarknoten. Für die ausgehende Explorer-Nachricht $msg_{out}$ gilt: $msg_{out}.treeParent = msg_{in}.parant$. Demnach kennt jeder Empfänger einer Explorer-Nachricht $msg_{in}$ einen $msg_{in}.parent$, den Absender der

**Algorithmus 2 :** Auszüge aus dem DIBADAWN Algorithmus als Pseudocode

   **Function** *recvForwardMessage(msg)*
      **if** *not visited* **then**
         visited = true;
         parent = msg.forwardedBy;
         startTimeout();
         myTTL = msg.ttl;
**1**         **if** *myTTL > 0* **then**
            msg.ttl = msg.ttl - 1;
            msg.forwardedBy = thisNode;
            msg.parent = parent;
            sendBroadcastWithJitteredDelay(msg)
      **else if** *msg.parent is not thisNode* **then**
         crossEdges.add(msg.forwardedBy)

   **Function** *onTimeout()*
      detectCycles();
      forwardMsg();
      detectArticulationPoints();
**2**      voteArticulationPoints();
**3**      voteBridges();

   **Function** *detectCycles()*
      **for** $m \in crossEdges$ **do**
         cycleId = createUniqueCycleId();
         edgeMarkings.add(time, m.id, NOBRIDGE, (thisNode, m.forwardedBy), 1);
         addPayloadToMsgFrom(m.forwardedBy, cycleId);
         bufferBackMsg(parent, cycleId);

   **Function** *recvBackMsg(m)*
      **if** $BRIDGE \in m.payload$ **then**
**4**         edgeMarking.add(time, m.id, BRIDGE, 1) `/* 1 := 100% competence */`;
         addPayloadToMsgFrom(m.forwardedBy, BRIDGE);
      **else**
         **for** $p \in m.payload$ **do**
            m2 = messageBuffer.getMsgContaining(p);
            **if** $m2 \neq null$ **then**
               messageBuffer.remove(m2);
               competence = calcCompetence(m);
**5**               edgeMarkings.add(time, m.fid, (thisNode, m.forwardedBy), competence);
               competence2 = calcCompetence(m2);
**6**               edgeMarkings.add(time, m2.fid, (thisNode, m2.forwardedBy), competence2);
            **else**
               messageBuffer.add(m);

empfangenen Explorer-Nachricht, und einen $msg_{in}.treeParent$, den ursprünglichen Absender einer Explorer-Nachricht. Das heißt, ein Empfänger empfängt einen Explorer vom Absender $msg_{in}.parent$ und dieser Absender hat diesen Explorer zuvor vom ursprünglichen Absender $msg_{in}.treeParent$ empfangen. Ein ursprünglicher Absender existiert nicht, wenn eine Explorer-Nachricht vom Initiator empfangen wird. Eine Verbindung zwischen einem Knoten und seinem parent entspricht genau einer Kante eines Spannbaumes mit dem Initiator als Wurzel. Mit dem wiederholten Weitersenden wird dieser Spannbaum schrittweise vervollständigt. Dieses Weitersenden des Explorers wird um eine zufällige Zeit verzögert, um die Häufigkeit von Kollisionen beim Weitersenden zu vermindern.

b  Der empfangene Explorer wird vom ursprünglichem Absender empfangen.
Das heißt, der ursprünglichem Absender, $msg_{in}.treeParent$, der Explorer-Nachricht entspricht dem Empfänger der Explorer-Nachricht. In diesem Fall wird die Explorer-Nachricht ignoriert, da sich die Suche in der Forward-Phase zirkulär ausgehend vom Initiator ausbreiten soll. Dieser Fall tritt sehr häufig ein, da der Empfänger einer Explorer Nachricht meist auch den Absender beim Weiterleiten des Explorers per Broadcast erreicht.

c  Der Empfänger-Knoten kennt den empfangenen Explorer, aber der Empfänger-Knoten entspricht nicht dem ursprünglichen Absender.
Das heißt, der ursprüngliche Absender, $msg_{in}.treeParent$, der Explorer-Nachricht entspricht nicht dem Empfänger der Explorer-Nachricht und der Explorer ist dem Empfänger dennoch bekannt. In diesem Fall wird ein Kreis im Graphen erkannt. Das Erkennen von Kreisen ist besonders wichtig. *Kreise*, also Graphen mit einem Pfad mit mindestens drei Knoten bei dem der Startknoten und der Endknoten miteinander verbunden sind, enthalten keine Brücken [Die00, S.11]. Zu jedem Pfad auf einem Kreis mit einer Richtung existiert ein alternativer Pfad mit der Gegenrichtung. Daher kann keine Kante auf einem Kreis eine Brücke sein.

Das Wissen über die Existenz eines Kreises wird vom Empfänger-Knoten in Form einer Querverbindung lokal vermerkt und der Explorer wird nach dem Erkennen des Kreises nicht weitergeleitet. Querverbindungen zwischen zwei Knoten in einer DIBADAWN-Ausführung sind stets eineindeutig, um sie im späteren Verlauf zweifelsfrei identifizieren zu können. Ein mögliches Vorgehen, um einen Kreis eineindeutig zu beschreiben, ist die Verbindung von Explorer, Absender und Empfänger. Hierfür könnte man den drei Eigenschaften Exploreridentität, Absenderidentität und Empfängeridentität jeweils einen numerischen Wert zuordnen und diese drei Werte der Größe nach sortiert miteinander verbinden und als eine Querverbindung speichern. Somit können zwei durch eine Querverbindung verbundene Knoten einen Kreis eineindeutig und einheitlich beschreiben, obgleich sie über vertauschte Paare von Absender und Empfänger verfügen.

Ein Kreis wird also erkannt, indem zwei Instanzen eines Explorers ausgehend von einem Knoten $v_{Start}$ zwei verschiedene Pfade eines Spannbaumes entlangwandern, bis sie erstmals auf zwei Knoten treffen, welche durch einen gemeinsamen

Link miteinander verbunden sind. Im DIBADAWN-Kontext wird dieser Link als Querverbindung betrachtet und repräsentiert den Kreis. Eine Besonderheit dieser Querverbindung ist ihre Position im Kreis. Die Querverbindung befindet sich "gegenüber" vom Knoten $v_{Start}$. Das bedeutet, dass die beiden an die Querverbindung angrenzenden Knoten ähnlich weit von $v_{Start}$ entfernt sind. Genauer gesagt, ist ihr Abstand zu $v_{Start}$ in Hops gemessen identisch, wenn der Kreis eine ungerade Anzahl an Kanten besitzt und unterscheidet sich um 1, wenn die Anzahl der Kanten im Kreis gerade ist.

**Asymmetrische Querverbindungen** In diesem Abschnitt wird allgemein von einer fehlerfreien Übertragung ausgegangen, um die Funktionsweise des Algorithmus DIBADAWN anschaulich darzustellen. An dieser Stelle soll jedoch eine Behandlung eines Fehlers beschrieben werden, welcher durch eine fehlerhafte Übertragung einer Explorer-Nachricht entsteht. Der zu behandelnde Fehler ist eine asymmetrische Querverbindung. Solch ein Fehler kann auftreten, wenn zwei Instanzen $x_1, x_2$ eines Explorers ausgehend von einem Knoten zwei verschiedene Pfade eines Spannbaumes entlangwandern, wobei $x_1$ fehlerhaft von einem Knoten $v_1$ übertragen und somit nicht weitergeleitet wird. Das Resultat ist, dass der Explorer $x_2$ immer weiter auf dem Kreis entlangwandert bis zu einem Knoten $v_2$, welcher die Explorerinstanz $x_2$ zu $v_1$ erfolgreich überträgt. Somit wird eine Querverbindung von $v_1$ erkannt, jedoch erkennt $v_2$ diese Querverbindung nicht, da $v_2$ nicht die Explorerinstanz $x_1$ empfing. Wie später beschrieben führt dies dazu, dass eine einzelne Kreis-Nachricht in der Backward-Phase bis zum Initiator übertragen wird und somit das Erkennen von Brücken auf dem gesamten Pfad im Spannbaum zwischen $v_1$ und dem Initiator verhindert. Auf eine von Knoten $v$ erkannte asymmetrische Querverbindung wird so reagiert, dass die Querverbindung von $v$ ignoriert wird und somit nicht zu Beginn der Backward-Phase zum parent weitergeleitet wird. Daher können alle Kanten zwischen $v$ und dem höchsten Knoten $v_h$ im Spannbaum, welcher sich noch auf dem zugehörigen Kreis befindet fehlerhaft als Brücken erkannt werden, aber auf der anderen Seite ist es wie gewollt auch möglich, dass Brücken zwischen $v_h$ und dem Initiator erkannt werden. Asymmetrische Querverbindungen erkennt man über die Anzahl der Hops, die eine Explorerinstanz zurückgelegt. Ist eine Explorerinstanz über $n$ Hops zu einem Knoten gewandert und empfängt dieser Knoten eine weitere Instanz des selben Explorers über $m$ Hops mit $|n - m| > 1$, so wird eine asymmetrische Querverbindung erkannt.

## 4.5 Backward-Phase

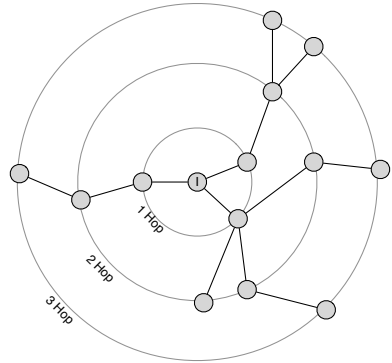

Abbildung 4.1: Funktionsweise des Timers

Nach dem Versenden einer Nachricht in der Forward-Phase aktiviert der Absender einen Timer. Der Ablauf eines Timers in einem Knoten bewirkt den Übergang des Knotens in die Backward-Phase. Die Bedeutung dieses Timers ist essentiell, er koordiniert die schrittweise Beendigung des DIBADAWN-Algorithmus. Der Grundgedanke ist, dass Timer eines Knoten mit einer größeren Entfernung zum Initiator im Spannbaum früher ablaufen, als ein Timer eines Knoten mit einer geringeren Entfernung zum Initiator. Dieser Zusammenhang ist in der Abbildung 4.1 dargestellt. Der Initiator I in der Mitte hat drei Nachbarknoten mit einer Entfernung von einem Hop. Diese Nachbarknoten haben wiederum Nachbarn mit einer größeren Entfernung, welche teilweise wieder Nachbarn mit einer noch größeren Entfernung zum Initiator haben. Somit ergibt sich in der Abbildung eine maximale Entfernung von drei Hops zum Initiator für die fünf Knoten auf dem äußersten Kreis. Daher sind die Timer so zu konfigurieren, dass zuerst alle Timer der Knoten auf dem äußersten Kreis ablaufen, danach die Timer der Knoten auf dem mittleren Kreis und schließlich die Timer der Knoten auf dem inneren Kreis ablaufen und abschließend der Timer des Initiator abläuft. Die verwendete Dauer für einen Timer ist abhängig von der Entfernung zum Initiator und der Anzahl der maximal vom Algorithmus zu erreichenden Knoten. Hierfür wird die Entfernung in Hops, also die Anzahl der Übertragungen, genutzt. Dabei ist die Dauer so zu wählen, dass jeder Timer abläuft bevor ein Timer mit einer geringeren Entfernung zum Initiator abläuft. Nach dem Ablauf eines Timers muss zusätzlich genügend Zeit vorhanden sein, um eine Nachricht vom zugehörigen parent versenden zu können.

Anders als in der Forward-Phase werden in der Backward-Phase Nachrichten ausschließlich per Unicast von einem Knoten zum zugehörigen parent, also in Richtung Wurzel des Spannbaumes, versendet. Da sich Knoten mit einer höheren Entfernung zum Initiator früher in der Backward-Phase befinden, werden Nachrichten aus der Backward-Phase bereits während der Forward-Phase vom zugehörigen parent empfangen. Es gibt zwei Arten von Nachrichten in der Backward-Phase: Brücken-Nachrichten und Kreis-Nachrichten. Eine Brücken-Nachricht beschreibt eine Brücke zwischen ihrem Absender und ihrem Empfänger. Eine Kreis-Nachricht beschreibt einen Kreis im Graphen eindeutig und der Empfang der Kreis-Nachricht bedeutet, dass der Link zwischen dem Absender und dem Empfänger der Kreis-Nachricht eine Kante des beschriebenen Kreises ist. Brücken-Nachrichten und durch Kreis-Nachrichten beschriebene Querverbindungen sind innerhalb einer DIBADAWN-Ausführung eineindeutig, also eine Querverbindung beschreibt genau einen Kreis und ein Kreis wird genau durch eine Querverbindung beschrieben. Eine Brücken-Nachricht beschreibt genau eine Brücke und eine Brücke wird genau durch eine Brücken-Nachricht beschrieben. Beim Empfang einer Nachricht aus der Backward-Phase sind folgende Fälle zu unterscheiden:

a  Eine Brücken-Nachricht wird empfangen.
   Dies bedeutet, dass der Absender die Existenz der Brücke erkannt hat und den Empfänger darüber informiert, sodass beide Knoten über diese Information lokal verfügen. Diese Nachricht wird nicht weitergeleitet und somit bleibt das Wissen über die erkannte Brücke lokal und steht nicht allgemein allen Knoten im Netzwerk zur Verfügung.

b  Eine Kreis-Nachricht wird empfangen und der von ihr beschriebene Kreis ist dem Empfänger nicht bekannt.
   Diese Nachricht wird vom Empfänger bis zum Ablauf des eigenen Timers in einem Nachrichtenausgangsspeicher abgelegt. Dabei ist es möglich, dass die Nachricht entweder, wie im nächsten Fall beschrieben, durch den Erhalt einer weiteren gleichen Kreis-Nachricht entfernt wird oder in der Backward-Phase zum parent versendet wird.

c  Eine Kreis-Nachricht wird empfangen und der von ihr beschriebene Kreis ist dem Empfänger bereits bekannt.
   Dies bedeutet, der Empfänger wurde von genau zwei unterschiedlichen Absendern darüber informiert, dass die beiden Links zwischen dem Empfänger und den zwei Absendern ein Bestandteil eines gemeinsamen Kreises sind. Dies ist genau dann der Fall, wenn der Empfänger der Kreis-Nachricht ein gemeinsamer Vorfahre der beiden Knoten ist, welche diesen Kreis zuerst in Form einer Querverbindung in der Forward-Phase erkannten. Das heißt, beide durch die Querverbindung verbundenen Knoten versenden jeweils eine Instanz der Kreis-Nachricht entlang zweier Pfade des Spannbaumes, bis sich diese Instanzen in einem gemeinsamen Knoten, dem gemeinsamen Vorgänger, treffen. Dieser gemeinsame Vorgänger, der Empfänger in diesem Fall, vermerkt sich die entsprechenden Links als Nicht-Brücken und entfernt alle Kreis-Nachrichten aus dem Nachrichtenausgangsspeicher, welche diesen Kreis beschreiben.

Geht ein Knoten in die Backward-Phase über, so erstellt er zuerst Kreis-Nachrichten für die von ihm erkannten Querverbindungen, also über die erkannten Kreise, und legt diese Kreis-Nachrichten im Nachrichtenausgangsspeicher ab. Anschließend wird der gesamte Inhalt seines Nachrichtenausgangsspeichers an seinen parent versendet. Um die Nutzung des Mediums zu minimieren werden die zu versenden Nachrichten nach Möglichkeit gemeinsam versendet. Danach wird geprüft, ob der Knoten selbst ein Gelenkpunkt ist. Abschließend wird die Statistikkomponente verwendet, um die gewonnenen Ergebnisse für eine lokale Nutzung zu verbessern.

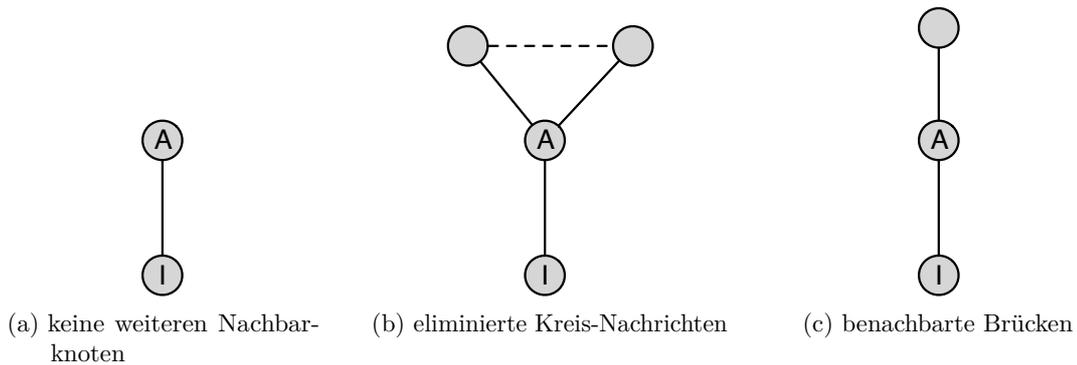

(a) keine weiteren Nachbarknoten  (b) eliminierte Kreis-Nachrichten  (c) benachbarte Brücken

Abbildung 4.2: Ursachen für einen leeren Nachrichtenausgangsspeicher

Nach dem Erstellen der Kreis-Nachrichten zu Beginn der Backward-Phase wird der gesamte Inhalt des Nachrichtenausgangsspeichers an den parent im Spannbaum versendet. Jedoch kommt es vor, dass der Nachrichtenausgangsspeicher leer ist. In diesem Fall wird eine Brücke zwischen dem Knoten und seinem parent lokal erkannt und diese Brücke wird mit einer Brücken-Nachricht an den zugehörigen parent im Spannbaum gesendet, sodass beide durch die Brücke verbundenen Knoten über die Existenz der Brücke informiert sind. Die Ursachen für einen leeren Nachrichtenausgangsspeicher sind vielfältig, lassen sich aber auf die in der Abbildung 4.2 dargestellten Basisfälle zurückführen. In den drei abgebildeten Basisfällen ist der Knoten I der Initiator und der Knoten A jeweils der Knoten mit dem leeren Nachrichtenausgangsspeicher. Im ersten Fall 4.2a verfügt der Knoten A ausschließlich über den Link zum Knoten I und somit empfängt A nur den Explorer von I und keine weiteren Nachrichten zur Erkennung von Kreisen bzw. Kreisnachrichten zu bereits erkannten Kreisen. Im zweiten Fall 4.2b verfügt der Knoten A neben dem Link zu I nur über Links, welche paarweise zu einem Kreis gehören und dessen Kreis-Nachrichten daher beim Eliminieren aus dem Nachrichtenausgangsspeicher entfernt werden, womit der Nachrichtenausgangsspeicher komplett geleert wird. Im dritten Fall 4.2c verfügt der Knoten A außer neben dem Links zum Knoten I nur über Links, welche Brücken sind, und Brücken-Nachrichten werden nicht weitergesendet und erscheinen daher auch nicht im Nachrichtenausgangsspeicher.

Nach dem Erstellen der Kreis-Nachrichten und dem Versenden des Nachrichtenausgangsspeichers zu Beginn der Backward-Phase prüft ein Knoten, ob er selbst ein Gelenkpunkt ist. Nach Definition 2.5 ist ein Knoten ein Gelenkpunkt, wenn er mehrere disjunkte Komponenten verbindet. Das heißt, es gilt zu prüfen, ob der eigene Knoten Bestandteil mehrerer Komponenten ist. Hierfür werden alle Ein- und Ausgänge von Brücken- und Kreis-Nachricht protokolliert. Das bedeutet, dass jedem Nachbarn eine Menge von gesendeten bzw. empfangenden Nachrichten zugeordnet wird. Anhand dieser Informationen wird ermittelt, welche Links zwischen dem eigenen Knoten und den Nachbarknoten sich in einer **Kreisrelation** befinden. Zwei Links befinden sich in

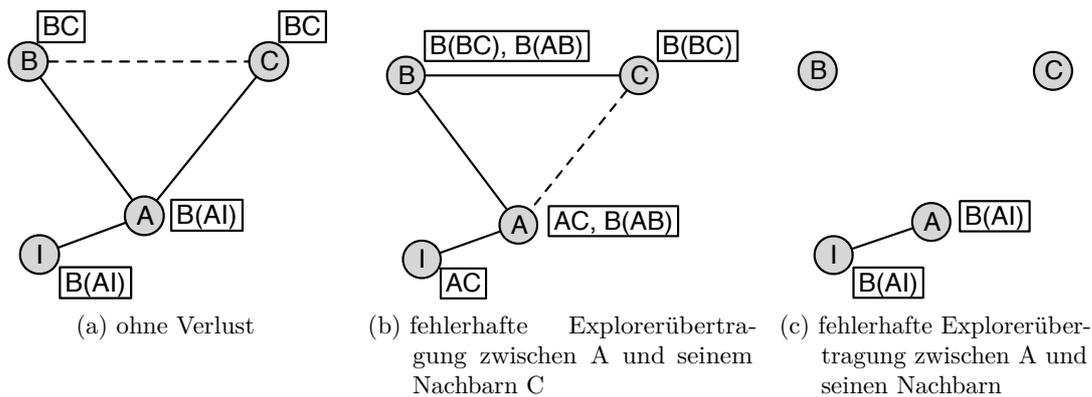

(a) ohne Verlust  (b) fehlerhafte Explorerübertragung zwischen A und seinem Nachbarn C  (c) fehlerhafte Explorerübertragung zwischen A und seinen Nachbarn einer Kreisrelation, wenn es einen gemeinsamen Kreis im Netzwerk gibt, auf welchem sich diese Links befinden. Aus der algorithmischen Sicht befinden sich diejenigen Links in Kreisrelation, welche über gemeinsame Nachrichten im Protokoll verfügen. Diese Kreisrelation ist transitiv, daher gilt, wenn sich die Links A und B in Kreisrelation befinden und sich die Links B und C in Kreisrelation befinden, dann befinden sich auch die Links A und C in Kreisrelation. Um zu prüfen, ob der eigene *Knoten* Bestandteil mehrere Komponenten ist, wird für die Links zu allen Nachbarknoten die transitive Hülle über der Kreisrelation ermittelt. Existiert ein Link zu einem Nachbarknoten, welcher nicht Bestandteil dieser transitiven Hülle ist, so verbindet der eigenen Knoten mehrere Komponenten und ist somit ein Gelenkpunkt.

Abbildung 4.3: Beispiel fehlerhafter Nachrichtenübertragung

Nach dem Erstellen der Kreis-Nachrichten, dem Versenden des Nachrichtenausgangsspeichers und der Gelenkpunkt-Prüfung befindet sich die DIBADAWN-Ausführung in einem Knoten bereits am Schluss. Dieser Knoten verfügt nun über die Aussagen, ob er ein Gelenkpunkt ist und welche seiner Links Brücken sind. Um die Zuverlässigkeit dieser Aussagen zu steigern wird abschließend die Statistikkomponente verwendet. Danach gilt die Suche nach Brücken und Gelenkpunkten als beendet.

## 4.6 Statistik-Komponente

In drahtlosen Netzwerken kann es zu fehlerhaften Übertragungen kommen. Hierdurch kann die Ausführung von DIBADAWN fehlerhafte Resultate hervorbringen. Der Anspruch, das Übertragungsmedium mit möglichst wenigen Nachrichten zu belasten bedeutet, dass der Verlust einer der wenigen Nachrichten drastische Auswirkungen für das Erkennen von Brücken und Gelenkpunkten hat.

**Beispiel einer fehlerhaften Übertragung**  In der Abbildung 4.3 sind zwei Beispiele einer fehlerhaften Übertragung dargestellt. In Abbildung 4.3a ist der Zustand nach eine DIBADAWN-Ausführung zu sehen. Ausgehend vom Initiator I wird ein Explorer über A an B und C erfolgreich versendet. Dies ist mit einer durchgehenden Geraden IA, AB und AC gekennzeichnet. Weiter wird die Querverbindung BC von den Knoten B und C erkannt. Die Querverbindung ist als gestrichelte Gerade BC dargestellt und in den rechteckigen Boxen bei den Knoten B und C wird die erkannte Querverbindung eingetragen. Beide Instanzen dieser Querverbindung werden an den gemeinsamen parent A gesendet und von diesem eliminiert. Weitere Querverbindungen sind A nicht bekannt und daher erkennt A die Brücke AI, dargestellt als B(AI), und sendet eine Brücken-Nachricht an seinen parent I. Abschließend stellt A noch fest, dass zwar B und C in Relation zueinander stehen aber I mit keinem Nachbarn in Relation steht und A somit ein Gelenkpunkt ist.

Wird anfangs der Explorer fehlerhaft von A zu C übertragen, so erhält man den Spannbaum aus Abbildung 4.3b und statt der Querverbindung BC wird die Querverbindung AC nur einmal von A erkannt. Wodurch C die Brücke BC erkennt und diese an B sendet. Der Knoten B sendet diese Brücke nicht weiter, erkennt die Brücke AB und sendet sie weiter an A. Da A die Querverbindung AC erkannt hat, sendet A diese Querverbindung weiter an I. Hierdurch wird die Brücke AI nicht erkannt. Weiterhin wird A als Gelenkpunkt erkannt, da schon I und B über keine gemeinsamen Nachrichten verfügen und somit keine Relation zwischen allen Knoten besteht.

Wird anfangs der Explorer fehlerhaft von A zu B und von A zu C übertragen, so erhält man den Graphen in der Abbildung 4.3c. Hier erkennt A die Brücke AI und sendet diese an den parent I. I empfängt die Brücken-Nachricht und die Ausführung des Algorithmus wird terminiert, ohne A als Gelenkpunkt zu erkennen. Die Knoten B und C bilden jeweils eine Komponente mit einem Knoten.

**Kompetenz**

Die Unzuverlässigkeit bei der Übertragung führt dazu, dass Brücken und Gelenkpunkte fehlerhaft erkannt werden. Das heißt, Knoten, welche keine Gelenkpunkte sind, werden als Gelenkpunkte erkannt und Gelenkpunkte werden nicht als solche erkannt. Kanten, welche keine Brücken sind, werden als Brücken erkannt und Brücken werden nicht als solche erkannt. Um dieser Unzuverlässigkeit entgegenzuwirken ist eine Statistik-Komponente in den DIBIDAWN-Algorithmus integriert.

Wird in der Backward-Phase eine Brücken-Nachricht empfangen, so wird davon ausgegangen, dass diese Information korrekt ist. Der Grund hierfür ist die geringe Reichweite dieser Nachricht. Eine Brücke wird von einem Knoten A in Form einer Querverbindung erkannt und mit einer Brücken-Nachricht an den parent von A weitergeleitet. Eine empfangene Brücken-Nachricht wird nicht weitergeleitet. Sie wird entweder vom parent empfangen oder gar nicht empfangen. Im Gegensatz wird beim Empfang einer Kreis-Nachricht im Rahmen einer Plausibilitätsprüfung eine Wahrscheinlichkeit für die Korrektheit dieser Aussage ermittelt. Der Grund hierfür ist wiederum die Reichweite dieser Nachricht. Kreis-Nachrichten werden entlang des Spannbaumes in Richtung Wurzel versendet und wandern solange einen Pfad entlang, bis zwei gleiche Kreis-Nachrichten in einem gemeinsamen Knoten eliminiert werden. Dies kann maxi-mal der gesamten Tiefe des Spannbaumes entsprechen. Die Wahrscheinlichkeit für die Korrektheit einer Nachricht wird als Kompetenz der Nachricht bezeichnet und sie wird berechnet über die Anzahl der Hops, die diese Nachricht bis zum Empfänger benötigt.

| Hops $h$ | $P(h)$    | Hops $h$ | $P(h)$    | Hops $h$ | $P(h)$    | Hops $h$ | $P(h)$    |
|----------|-----------|----------|-----------|----------|-----------|----------|-----------|
| 1        | 0.95      | 7        | 0.9899482 | 13       | 0.9666267 | 19       | 0.8891968 |
| 2        | 0.90      | 8        | 0.9877227 | 14       | 0.9592378 | 20       | 0.8646647 |
| 3        | 0.9954834 | 9        | 0.9850044 | 15       | 0.9502129 | >20      | 0.85      |
| 4        | 0.9944834 | 10       | 0.9816844 | 16       | 0.9391899 |          |           |
| 5        | 0.9932621 | 11       | 0.9776292 | 17       | 0.9257264 |          |           |
| 6        | 0.9917703 | 12       | 0.9726763 | 18       | 0.9092820 |          |           |

Tabelle 4.6.1: Wahrscheinlichkeit für eine korrekte Übertragung einer Kreis-Nachricht

Die hierfür verwendete Abbildung, dargestellt in der Tablle 4.6.1, wurde empirisch durch Simulationen erhalten[Mil09, S. 100]. Die konkreten Werte dienen hierbei weniger dem Verständnis des Algorithmus, sondern eher der Replizierbarkeit der Versuche im Abschnitt 7.

**Voting**

An der aktuellen Stelle ist der Algorithmus nun fast vollständig beendet. Da aus den oben genannten Gründen davon auszugehen ist, dass es zu fehlerhaften Aussagen bei der Brücken- und Gelenkpunktsuche bei der aktuellen Ausführung kam, wird versucht, die aktuellen Aussagen mit vergangenen Aussagen zu stützen. Daher wird im Algorithmus 2 in der Funktion onTimeout in den Zeilen 2 und 3 ein Mechanismus zur Stabilisierung der Aussagen ausgeführt.

Während jeder Ausführung von DIBADAWN trifft ein Knoten Aussagen darüber, ob eine Kante entweder eine Brücke- oder keine Brücke ist und der Knoten selbst entweder ein Gelenkpunkt- oder kein Gelenkpunkt ist. Diese Aussagen werden während der Ausführung (4.3) verarbeitet und anschließend lokal vom Knoten gespeichert. Diese gespeicherten Aussagen dienen der Statistikkomponente als Grundlage für eine Stabilisierung für zukünftige Aussagen über Kanten und den eigenen Knoten. Die Statistikkomponente ist somit in der Lage, anhand der vergangenen $k$ Aussagen zuverlässigere Entscheidungen zu treffen. Diese Entscheidungen sind als die eigentlichen Resultate des DIBADAWN-Algorithmus zu verstehen und können von anderen Anwendungen abgefragt werden. Im Gegensatz hierzu sind die Aussagen einer DIBADAWN-Ausführung als ein Zwischenergebnis zu verstehen.

Die Art, wie die letzten $k$ Aussagen für die Stabilisierung der entgültigen Aussage verwendet werden, wird mit der Wahl einer entsprechenden Regel bestimmt [Mil09, S. 99]. Hierfür stehen die folgende Regeln zur Auswahl, welche nachfolgend exemplarisch für die Brückenerkennung beschrieben werden.

1. Einstimmigkeits-Regel

2. Einfache-Mehrheits-Regel

3. Eine-Stimme-Regel

4. Itelligent-Mehrheits-Regel

5. Kompetenter-Kreis-Regel

6. Gewichtete-Regel

**Einstimmigkeits-Regel**  Bei dieser Regel wird eine Kante als Brücke erkannt genau dann, wenn die letzten $k$ Aussagen diese Kante ebenfalls als Brücke bestätigen und somit einstimmig darüber entscheiden, dass diese Kante eine Brücke ist. Sobald eine Kante nur einmal als Nicht-Brücke erkannt wurde, so wird diese Kante mindestens für die nächsten $k$-Mal als Nicht-Brücke erkannt. Wurde in der Vergangenheit noch keine Aussage über eine Kante getroffen, so gilt die Kante einstimmig als Brücke, da Null von Null Stimmen dies bestätigen.

**Einfache-Mehrheits-Regel**  Bei dieser Regel wird eine Kante als Brücke erkannt genau dann, wenn mehr als die Hälfte der letzen $k$ Aussagen ebenfalls diese Kante als Brücke erkannten. Somit kann eine einmalig fehlerhafte Aussage korrigiert werden.

**Eine-Stimme-Regel**  Bei dieser Regel wird eine Kante als Brücke erkannt, sobald nur eine der letzten $k$ Aussagen diese Aussage unterstützt. Somit wird immer eine Brücke erkannt, wenn nicht einstimmig keine Brücke erkannt wird.

**Intelligente-Mehrheits-Regel**  Bei dieser Regel wird eine Kante genau dann als Brücke erkannt, wenn mehr als die Hälfte der letzen $k$ Aussagen ebenfalls diese Kante als Brücke erkannten und zusätzlich keine vertrauenswürdige DIBADAWN-Nachricht empfangen wurde, welche diese Kante als Nicht-Brücke bezeichnet.

**Kompetenter-Kreis-Regel**  Bei dieser Regel wird eine Kante genau dann als Brücke erkannt, wenn keine vertrauenswürdige DIBADAWN-Nachricht empfangen wurde, welche diese Kante als Nicht-Brücke bezeichnet.

**Gewichtete-Regel**  Bei dieser Regel wird eine Kante genau dann als Brücke erkannt, wenn die Summe der letzen gewichteten $k$ Aussagen über einem Schwellwert des Netzwerkes liegt. Wobei die Gewichte für Aussagen über eine Brücke positiv und Aussagen über eine Nicht-Bücke negativ in die Summe eingehen. Hierfür wird jeder Kompetenz einer DIBADAWN-Nachricht ein Gewicht mit der nachfolgenden Abbildung *weight* zugeordnet. Der Schwellwert des Netzwerkes wird erhalten, indem ein Gewicht für die Wahrscheinlichkeit einer Brücke im Netzwerk ebenfalls mit der Abbildung *weight* bestimmt wird. Die Wahrscheinlichkeit einer Brücke in einem Netzwerk ergibt sich aus der Anzahl aller Kanten $N_T$ und der Anzahl der Brücken $N_B$ in diesem Netzwerk mit $N_B/N_T$.

$$\begin{aligned} weight : p &\longrightarrow w \\ p &\mapsto log_e\left(\frac{p}{1-p}\right) \end{aligned}$$

# 5 Implementation

Dieser Abschnitt soll auf der einen Seite einen Überblick über die Softwarearchitektur der Implementierung geben und den Entwicklungsprozess umreißen ohne zu sehr ins Detail zu gehen. Auf der anderen Seite soll in diesem Abschnitt der Fokus auf den Problemen und verwendete Problemlösungen liegen, welche bei der Entwicklung aufgetreten sind. Abschließend wird ein Ausblick für mögliche zukünftige Erweiterungen gegeben, welche die Nutzung der Software durch einen Dritten erleichtern könnte.

## 5.1 Softwarearchitektur

Im Rahmen dieser Arbeit wurde eine Software entwickelt, welche die Funktionalität des Algorithmus DIBADAWN beinhaltet. Es wurde ein Modul in der objektorientierten Programmiersprache C++ für das Click Modular Router Framework entwickelt und anschließend evaluiert.

Das entwickelte Softwaremodul ist eines von weiteren Softwaremodulen, welche bisher im Rahmen des Berlin Roof Net Projekt entwickelt wurden. Das Berlin Roof Net ist ein Projekt von Studenten der Humboldt Universität zu Berlin mit der Motivation, die drahtlose Internetanbindung eines urbanen Bereiches zu verbessern. Dabei soll das Netzwerk vollständig selbstorganisierend und dezentral strukturiert sein. Somit soll keine permanente Verbindungstopologie existieren und keine Konfiguration jedes einzelnen Knotens im Netzwerk erforderlich sein. Das heißt, Netzwerkteilnehmer können sich dem Netzwerk anschließen, aus diesem austreten und sich in diesem bewegen, woraufhin diese Änderungen vom Netzwerk erkannt werden. Die Reaktion auf derartige Änderungen soll begrenzt sein, um den netzeigenen administrativen Aufwand gering zu halten und somit ausreichend Kapazitäten für Nutzdaten zu bewahren [Roo13].

**Click Modular Router**

Die Softwarearchitektur Berlin Roof Net Projektes ist so konzipiert, dass die einzelnen algorithmischen Ideen in Form von Elementen für einen Click Modular Router implementiert werden. Der Click Modular Router, entwickelt am Massachusetts Institute of Technology in Zusammenarbeit mit der University of California in Los Angeles, ist eine Softwarearchitektur für die Entwicklung von Routern. Hierbei wird ein Router als eine Liste von Elementen verstanden. Diese Elemente arbeiten paketorientiert und übernehmen Teilaufgaben eines Routers. Entsprechend ihrer Position in der Liste sind Elemente so miteinander verknüpft, dass die gewünschte Funktionalität des Routers erhalten wird. Derartige algorithmische Teilaufgaben reichen von einer einfachen Warteschlangenfunktionalität bis zu komplexen Protokollen [KMC+00].

**NS2 Simulator**

Für die Entwicklung und das Testen von Softwaremodulen wird im Berlin Roof Net Projekt der Simulator "Network Simulator" in der Version 2, kurz NS2, verwendet. Der NS2 Simulator wird seit 1989 als Open-Source-Projekt von verschiedenen Institutionen aus dem Umfeld der Forschungsgemeinde für Netzwerkkommunikation entwickelt. Unter ihnen die Cornell University, Defense Advanced Research Projects Agency, kurz DARPA und die National Science Foundation. Der NS2 Simulator ist forschungsorientiert konzipiert. Mit ihm lassen sich kabelgebundene und drahtlose Kommunikationsnetzwerke simulieren. Die Arbeitsweise des NS2 Simulators ist ereignisorientiert, das heißt die von dem Simulator simulierten Ereignisse werden zu nicht-äquidistanten Zeitpunkten bearbeitet [IH08]. Seit 2006 existiert der Simulator NS3, welcher als Nachfolger des Simulators NS2 ohne Abwärtskompatibilität vorgesehen ist. Der NS3 Simulator gleicht diverse Schwächen des NS2 Simulator aus, darunter die fehlende Parallelisierbarkeit bei der Ausführung und die wenig bekannte Programmiersprache oTCL wird nicht mehr als Konfigurationsschnittstelle verwendet [KBO13]. Auf einen Wechsel von NS2 zum NS3 Simulator wird derzeit verzichtet, da die Portierung der bisher implementierten Software einen zu großen Aufwand darstellt.

Es werden also der NS2 Simulator und der Click Modular Router verwendet. Diese Kombination wird als "nsclick" bezeichnet und bietet den Vorteil, dass implementierte Algorithmen nicht erneut für ein mögliches Zielsystem implementiert werden müssen. Eine Implementation eines Algorithmus verwendet das Click Modular Router Framework und wird entweder für eine Ausführung im NS2 Simulator oder für eine Ausführung auf einem realen Zielsystem übersetzt [NJG02]. Somit werden mögliche Fehler einer erneuten Implementierung, welche im Softwareentwicklungsprozess auftreten können, vermieden. Zugleich wird die Nutzung von Optimierungsmöglichkeiten des Compiler ermöglicht, welcher für das Zielsystem entwickelt wurde.

Das im Rahmen dieser Arbeit entwickelte Element für den Click Modular Router heißt TopologyDetection. Über diesen Namen kann eine zu verwendende Instanz für den Click Modular Router konfiguriert werden, um somit die Funktionalität das Algorithmus DIBADAWN zu nutzen. Das Element bietet eine Schnittstelle für die Konfiguration und Steuerung der Instanz sowie eine Schnittstelle zum Auslesen der Ergebnisse der DIBADAWN Ausführung. Da DIBADAWN ein dezentraler Algorithmus ist, verhält sich jede Instanz des TopologyDetection Elementes gleich und muss auf allen Netzwerkknoten aktiv sein. Weil DIBADAWN ein proaktiver Algorithmus ist, muss er regelmäßig ausgeführt werden, bevor ein anderes Element über die Ergebnisse von DIBADAWN verfügen kann. Über die Ergebnisse einer DIBADAWN-Ausführung können andere Elemente verfügen, indem sie eine Instanz des Elementes TopologyInfo mit DIBADAWN gemeinsam nutzen. Eine TopologyDetection-Instanz aktualisiert die Einträge in einer TopologyInfo-Instanz und diese Einträge werden wiederum von weiteren Instanzen anderer Elemente genutzt.

## 5.2 Test der Software

Bei der Implementierung des Softwaremoduls wurde nach dem iterativ-inkrementellen Wasserfallmodell vorgegangen. Die hierfür benötigten Tests wurden in Form von Simulations-Szenarios mit einer anschließenden automatisierten Auswertung der Simulationsergebnisse erstellt. So wurden beispielsweise drahtlose Maschennetzwerke mit einer Brücke, keiner Brücke, wenigen Brücken und vielen Brücken konstruiert, um das Erkennen von Brücken zu testen. Ähnliche Tests wurden ebenfalls für weitere Anwendungsfälle erzeugt.

## 5.3 Probleme und Lösungen

In diesem Abschnitt wird auf Probleme bei der Implementierung des Algorithmus DIBADAWN hingewiesen und mögliche Lösungsansätze für diese Probleme werden aufzeigt.

**Jitter** Bei der Weiterleitung von Informationen in einem drahtlosen Maschennetzwerk kann es zu Kollisionen kommen, da ein gemeinsames Übertragungsmedium von allen Netzwerkteilnehmern genutzt wird. Aufgrund von Kollisionen kann der Empfang bei einem oder mehreren Empfängern fehlerhaft verlaufen. Dies ist besonders kritisch bei der Nutzung von Broadcast-Nachrichten, da der Empfang dieser Nachrichten nicht bestätigt werden muss und eine Kollision somit unentdeckt bleibt. Um Kollisionen zu reduzieren, wird eine zufällige Verzögerungsdauer, auch Jitter genannt, beim Versenden einer Nachrichten verwendet. Da im Fall einer Kollision in der Forward-Phase ein oder mehrere Knoten nicht oder auf einem anderen Weg erreicht werden können, wird während einer DIBADAWN-Ausführung eine andere Topologie erkannt und möglicherweise werden aufgrund dieser erkannten Topologie andere Brücken und Gelenkpunkte erkannt.

In der ursprünglichen Beschreibung des DIBADAWN-Algorithmus wird zwar auf die Verwendung einer zufälligen Verzögerung hingewiesen aber eine detaillierte Beschreibung dieser Verzögerung bleibt leider aus. Immerhin wird mit der Darstellung der Verzögerung als $U(0, jitterMax)$ ein Hinweis auf die untere Grenze der Verzögerung gegeben [Mil09, S. 77]. Die untere Grenze wird als 0 beschrieben. Dies bedeutet, dass in diesem Fall effektiv keine Verzögerung statt findet. Die obere Grenze wird mit jitterMax angegeben, wie hoch diese ist und wie dieser Wert zu bestimmen ist, wird allerdings nicht beschrieben.

Daher wird in dieser Arbeit der Jitter nach der RFC 5148 bestimmt [CAD08]. Das heißt, der Jitter wird gleich verteilt erzeugt zwischen der unteren Grenze 0 und der oberen Grenze jitterMax. Um eine konkreten Wert für jitterMax zu finden, wird ein Intervall benötigt, in welchem Nachrichten versendet werden [CAD08, S. 7]. Ein konkreter Wert für dieses Sendeintervall ist konfigurierbar und wurde im Rahmen dieser Arbeit auf 40ms, der maximalen Übertragungsdauer mit Jitter [Mil09, S. 79], gesetzt.

Gemessen an diesem Sendeintervall $I_S$, darf jitterMax nicht größer als $I_S/2$ sein und sollte nicht kleiner als $I_S/4$ sein.

**Kompetenzbewertung** In der ursprünglichen Beschreibung des Algorithmus DIBA-DAWN ist ein Fehler enthalten. Im Pseudocode in der Funktion $receive\_backmessage(...)$ wird die Kompetenz für eine empfangene Brücken-Nachricht berechnet und die Kompetenz einer Kreis-Nachricht auf Eins, also absolut vertrauenswürdig, gesetzt, wenn zuvor das Gegenstück dieser Kreis-Nachricht bereits empfangen wurde [Mil09, S. 78]. Kreis-Nachrichten, welche nicht lokal terminiert wurden, werden weitergeleitet, indem sie in den $messageBuffer$ eingereiht werden. Brücken-Nachrichten werden nicht in den $messageBuffer$ eingereiht und auch nicht weitergeleitet. Das heißt, Brücken-Nachrichten werden stets mit einer 1-Hop Entfernung zum ursprünglichen Absender empfangen. Kreis-Nachrichten werden stets mit einer $n$-Hops Entfernung zum ursprünglichen Absender empfangen mit $n \geq 1$. Die Berechnung der Kompetenz für die Brücken-Nachrichten wird mit einer Abbildung realisiert, welche die Kompetenz einer Nachricht in Abhängigkeit von der Entfernung, in Hops, zum ursprünglichen Absender berechnet. Dabei fällt die Kompetenz umso kleiner aus, desto größer die Entfernung ist. Weil die Entfernung zum ursprünglichen Absender einer Brücken-Nachricht stets 1-Hop beträgt, ist die Kompetenz einer Brücken-Nachricht konstant [Mil09, S. 98 ff.].

Der eigentliche Grundgedanke der Kompetenz ist die Bevorzugung der Entscheidungen von Knoten mit einer geringeren Entfernung. Die Kreis-Nachrichten können über weite Distanzen im Netzwerk übertragen werden und mit der beschriebenen Abbildung wird den Entscheidungen mit geringerer Entfernung eine höhere Kompetenz bescheinigt. Da Brücken lediglich von den angrenzenden Knoten erkannt werden und somit das Wissen über eine Brücke in Form einer Brücken-Nachricht nicht im Netzwerk über die angrenzenden Knoten hinaus weitergeleitet wird, ist eine Bevorzugung näherer Brücken-Nachrichten nicht zielführend.

Daher wird im Rahmen dieser Arbeit davon ausgegangen, dass den Brücken-Nachrichten eine konstant hohe Kompetenz bescheinigt werden soll und für die Kreis-Nachrichten soll eine Kompetenz in Abhängigkeit von der Entfernung in Hops berechnet werden, wie in Algorithmus 2 in den Zeilen 4, 5 und 6 dargestellt. Die hierfür verwendete Abbildung ist in der Tabelle 4.6.1 zu finden.

**Häufige Kollisionen in der Backward-Phase** Aufgrund der Verwendung des iterativ-inkrementellen Wasserfallmodells wurde bei der Entwicklung erkannt, dass es beim Versenden der Nachrichten in der Backward-Phase häufig zu Kollisionen kam. Die Ursache hierfür ist die ursprüngliche Beschreibung des DIBADAWN-Algorithmus. Der Timer für den Übergang von der Forward-Phase in die Backward-Phase unmittelbar nach dem Eintreffen eines unbekannten Explorers aktiviert, wie in der ursprünglichen Funktion $receive\_forward\_message(...)$ zu sehen [Mil09, S. 77]. Beim Übergang in die Backward-Phase aufgrund des Timers in der Funktion $on\_timeout()$ werden Kreise versucht zu finden und anschließend wird eine Antwort-Nachricht zusammengestellt und an den parent gesendet, wie in der Funktion $forward\_msg()$ beschrieben [Mil09,

S. 78].

Da der Explorer in der Forward-Phase per Broadcast übertragen wird, ist davon auszugehen, dass alle benachbarte Knoten diesen Explorer zeitgleich oder zeitnah empfangen. Dementsprechend wird auch der Timer von allen benachbarten Knoten, welche diesem Explorer zuvor unbekannt war, zur gleichen Zeit aktiviert. Löst der Timer aus, wird die beschriebene Funktion *onTimeout*() von all diesen Knoten zur gleichen Zeit gerufen und die Antwort-Nachrichten werden zur gleichen Zeit zum parent versendet. Dies führt dann zur Kollision dieser Nachrichten.

Eine Rücksprache mit dem Autor hat ergeben, dass die ursprüngliche Beschreibung nicht vollständig ist und das Versenden der Antwort-Nachrichten zum parent zufällig verzögert werden solle. Dies wurde so implementiert und ist in der evaluierten Version enthalten.

**Häufige asymmetrische Kreise**  Beim Testen der Software in größeren drahtlosen Maschennetzwerken fiel eine erhöhte Anzahl asymmetrische Kreise auf. Der Grund hierfür ist die Berechnung der zufälligen Verzögerung für die Weiterleitung des Explorers. Die Verzögerung wird bestimmt mit $U(0, jitterMax)$ und besitzt einen Wert zwischen 0 und $jitterMax$ [Mil09, S. 77]. Im Versuch trat allerdings häufig der Fall auf, dass die Summe mehrerer Verzögerung klein genug war, um einen Explorer viel weiter entlang einer Seite eines Kreises zu übertragen, als es auf der anderen Seite geschah, wodurch asymmetrische Kreise erkannt wurden, ohne dass ein Nachrichtenverlust auftrat.

Auch an dieser Stelle hat eine Rücksprache ergeben, dass die die ursprüngliche Beschreibung nicht vollständig ist, da eine Weiterleitung in Hop-Gruppen angedacht war. Dabei besteht eine Hop-Gruppe $n$ aus allen Knoten, welche einen gemeinsamen Explorer über die gleiche Entfernung $n$, also mit $n$-Hops, vom ursprünglichem Absender empfangen haben. Jede dieser Gruppen leitet den Explorer nacheinander weiter, wobei eine Gruppe mit einer geringeren Entfernung stets zuerst den Explorer weiterleitet. Somit wird es ausgeschlossen, dass ein Knoten A mit einer geringeren Entfernung zum ursprünglichen Absender einen Explorer von einem anderen Knoten mit einer größeren Entfernung zum ursprünglichem Absender erhält, obwohl A seinen Explorer noch nicht weitergeleitet hat.

Dies ist realisiert mit einer dynamischen minimalen Verzögerung der Weiterleitung nach dem Empfang eines neuen Explorers. Um einen Explorer in der Forward-Phase weiterzuleiten ist eine konstante Hop-Dauer vorgesehen. Während dieser Hop-Dauer wird der Explorer von allen Knoten aus der zugehörigen Hop-Gruppe übertragen. Diese Hop-Dauer setzt sich, wie in Abbildung 5.1 als MaxTraversalTime dargestellt, zusammen aus dem Forward-Jitter, dem Min-Delay und der Sendedauer MaxTxTime. Jeder Knoten aus einer Hop-Gruppe verzögert das Weiterleiten um eine zufällig gewählte Dauer, dies entspricht dem Forward-Jitter. Mit dem Empfang eines Explorers wird dem Empfänger die zuvor verwendete Wartezeit Forward-Jitter des Absenders mitgeteilt und der Empfänger muss nun mindestens so lange warten, bis eine Hop-Dauer vollständig erreicht ist, um somit allen weiteren Knoten seiner Hop-Gruppe den Empfang eines Explorers zu ermöglichen. Diese minimale Wartezeit entspricht dem Min-Delay, wie in der Abbildung 5.1 dargestellt. Sie wird berechnet mit

$$\text{Min-Delay} = \text{MaxTraversalTime} - \text{Forward-Jitter} - \text{MaxTxTime}$$

Die effektive Wartezeit eines Knotens in der Forward-Phase entspricht somit der Summe von Min-Delay und Forward-Jitter. Zu beachten ist, dass die eigentliche Sendedauer, also die Mediumzeit, welche für die Übertragung der Explorer-Nachricht benötigt wird, nicht exakt bestimmt wird, sondern lediglich durch einen konstanten maximalen Wert beschrieben wird.

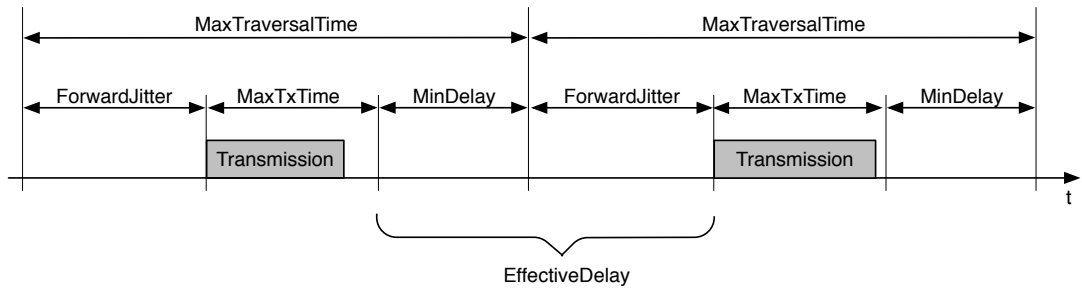

Abbildung 5.1: Verzögerung bei der Weiterleitung

**Entscheidungen für schlechte Links werden selten aktualisiert**  Bei der Evaluation fielen fehlerhafte Brücken auf, welche besonders oft ausgegeben aber nur sehr selten erkannt wurden. Die Ursache hierfür ist, dass Entscheidungen für manche Links seltener aktualisiert werden als für andere. Bei diesen Links handelt es sich um Links mit einer geringen Übertragungswahrscheinlichkeit. Daher werden Nachrichten über diese Verbindung auch nur selten erfolgreich übertragen und noch seltener erfolgreich hin- und zurückübertragen, wie es vom DIBADAWN-Algorithmus vorgesehen ist.

Die Ursache hierfür ist, dass in der ursprünglichen Beschreibung des DIBADAWN-Algorithmus über einen Link nur genau dann entschieden wird, sobald eine Brücke oder ein Kreis über diesen Link erkannt wird. In diesem Fall wird die Entscheidung in *edgeMarkings* gespeichert. Diese Entscheidung wird ausschließlich durch eine weitere Entscheidung für diesen Link aktualisiert. Genau hier liegt das Problem. Was geschieht, wenn der Link niemals aktualisiert wird? Beispielsweise könnte einer der beteiligten Knoten, welche mit einer Brücke verbunden sind, aus dem drahtlosen Maschennetz ausscheiden. Der hinterbliebene Knoten würde weiterhin den Inhalt von *edgeMarkings* ausgeben und somit die Existenz dieser Brücke behaupten.

Um dieses Problem zu lösen, werden die während einer Suche erkannten Brücken und Kreise in einer Menge zusammengefasst. Nach einer Suche wird diese Menge mit den letzten $n$ Mengen der vorherigen Suchen verglichen. Dabei werden fehlende Entscheidungen als Kreise, also Nicht-Brücken, betrachtet. Da Nicht-Brücken vom DIBADAWN-Algorithmus nicht ausgegeben werden, entsteht hierdurch keine Verfäl-schung der Ergebnismenge. Aber dies führt dazu, dass veraltete Entscheidungen nach und nach "vergessen" werden, weil sie von neueren Entscheidungen verdrängt werden.

## 5.4 Ausblick

Eine Verbreitung der lokal gefundenen Brücken und Gelenkpunkte ist nicht im DIBADAWN Algorithmus vorgesehen und wurde daher auch nicht implementiert. Das heißt, alle Instanzen von Click Modular Router Elementen auf einem Netzwerkknoten A können erfahren, ob A ein Gelenkpunkt im Netzwerk ist und A einen Link besitzt, welcher eine Brücke ist. Würde ein Element daran interessiert sein, alle Brücken oder alle Gelenkpunkte im Netzwerk zu kennen, so müsste ein Element die Verbreitung dieser Informationen selbst implementieren.

Die in der Forward-Phase von DIBADAWN verwendete Breitensuche wird ebenfalls von anderen Protokollen verwendet. Daher bietet es sich an, derartige Netzwerknachrichten zu bündeln und gemeinsam zu versenden, um die Anzahl der Nachrichten auf dem Medium und die Mediumzeit zu verringern.

Bei dem Design des Paketaufbaus wird darauf Wert gelegt, dass sich der Paketaufbau in Zukunft ändern kann und ältere Implementationen dies auch erkennen können. Der Grund hierfür ist die bisherige mangelnde Optimierung des Paketaufbaus. Die Pakete für den Nachrichtenaustausch zwischen den Netzwerkknoten sind teilweise ASCII-codiert, um eine Auswertung des Netzwerkverkehrs zu erleichtern. Solch ein Paketaufbau lässt sich bezüglich der Länge vermutlich optimieren und dies würde ebenfalls die Mediumzeit senken.

# 6 Methodik der Evaluation

In diesem Abschnitt der Arbeit wird die verwendete Methodik beschrieben und erläutert. Ebenso werden grundlegende Fragen diskutiert und für die spätere Verwendung in nachfolgenden Abschnitten festgehalten.

Das generelle Vorgehen bei der Evaluation des Algorithmus DIBADAWN ist folgendes. Anhand der tatsächlichen Verbindungen zwischen den Netzwerkteilnehmern eines drahtlosen Maschennetzwerkes wird ein Referenzgraph erstellt. In diesem Graphen werden die Brücken und Gelenkpunkte ermittelt und mit den Resultaten eines Versuchs mit mehreren DIBADAWN-Ausführungen verglichen. Der Vergleich mit mehreren DIBADAWN-Ausführungen ist notwendig, da der Algorithmus verteilt arbeitet und zusätzlich eine statistische Komponente enthält und diese Aspekte sonst nicht berücksichtigt werden.

Anders als bei kabelgebundenen Netzwerken sind die tatsächlichen Verbindungen zwischen den Netzwerkteilnehmern in einem drahtlosen Maschennetzwerk nicht offensichtlich und müssen ermittelt werden. Die reine geographische Anordnung der Netzwerkteilnehmer gibt kaum Aufschluss über die tatsächlich existierenden Verbindungen zwischen ihnen. Selbst Netzwerkteilnehmer, welche sich unmittelbar nebeneinander befinden, müssen nicht zwangsläufig miteinander verbunden sein. Beispielsweise können sich Barrieren, wie Pflanzen oder massive Wände, negativ auf die Kommunikation auswirken. Daher wird die Frage nach den tatsächlichen Verbindungen in einem drahtlosen Maschennetzwerk im Abschnitt 6.1 untersucht.

Ist der Referenzgraph gefunden, werden die Brücken und Gelenkpunkte in diesem Graphen ermittelt und mit den Ausführungen von DIBADAWN verglichen. Das Vergleichen und Bewerten der Ergebnisse wird im Abschnitt 6.3 beschrieben.

## 6.1 Was ist ein Link?

Möchte man eine Verbindungstopologie für ein drahtlosen Maschennetzwerk erhalten, so benötigt man neben der Anordnung der Teilnehmer noch die Information darüber, welche Teilnehmer mit welchen Teilnehmern verbunden sind.

Laut der Definition 2.8 besteht ein Link zwischen zwei Netzwerkteilnehmern genau dann, wenn Daten zwischen ihnen in eine oder beide Richtungen ausgetauscht werden können. Diese Definition ist zwar ausreichend für viele Anwendungsfälle, aber nicht praktikabel für drahtlose Maschennetzwerke. Unter welchen Umständen können Daten zwischen zwei Netzwerkteilnehmern ausgetauscht werden? Daten können ausgetauscht werden, wenn ein Netzwerkteilnehmer als potentieller Empfänger die Möglichkeit besitzt, Daten von einem Netzwerkteilnehmer, dem Absender, zu empfangen. In einem drahtgebundenen Netzwerk ist dies von der Verkabelung abhängig.

In einem gemeinsamen drahtlosen Maschennetzwerk trifft diese Definition allerdings für alle Paare von Netzwerkteilnehmern zu. Die Unterstützung für drahtlose Maschennetzwerke wurde mit dem Standard 802.11s™ in der MAC-Schicht, also in einer Teilschicht der Data-Link-Schicht des OSI-Modells, angesiedelt. Dies lässt die Betrachtung zu, dass ein Link zwischen zwei Netzwerkteilnehmern existiert ohne, dass eine direkte physikalische Verbindung zwischen ihnen vorhanden ist, welche diesen Datenaustausch mittels Radio-wellen nach dem Standard 802.11™ realisiert. Daher werden drahtlose Maschennetzwerke auch Multihop-Netzwerke genannt.

Die Daten würden hierfür dem Standard 802.11™ entsprechend über eine Liste $L$ von Netzwerkteilnehmern vom Absender zum Empfänger transportiert werden. Die Länge dieser Liste, einschließlich dem Absender und dem Empfänger, wird als $\#L$ notiert und entspricht der Anzahl der benötigten Verbindungen minus eins, welche nötig sind, um genau den Link zwischen dem ursprünglichen Absender und dem letzen Empfänger bereitzustellen. Diese hierfür benötigten Verbindungen werden Hops genannt.

**Beispiel** Sei in einem drahtlosen Maschennetzwerk mit den Netzwerkteilnehmern A,B und C. Ein Link zwischen A und C sei über B realisiert. Die entsprechende Liste ist demnach (A,B,C) und C ist über $h$ Hops von A erreichbar und es gilt folgende Gleichung.
$$h = \#(A,B,C) - 1 = 3 - 1 = 2$$

Unter einer n-Hopnachbarschaft eines Netzwerkteilnehmers X versteht man eine Menge an Netzwerkteilnehmern, welche genau n Hops von X entfernt sind. Im Beispiel ist die 1-Hopnachbarschaft von B die Menge {A,C} und die 1-Hopnachbarschaft von C die Menge {B}.

Da Multihopverbindungen ein inhärenter Bestandteil von drahtlosen Maschennetzwerken nach 802.11™ sind, soll an dieser Stelle die bisherige Definition eines Links (2.8) erweitert werden und hierfür werden folgende mögliche Erweiterungen für eine Verbindung zwischen zwei Teilnehmern in einem drahtlosen Maschennetzwerk betrachtet.

(a) Ein Link $L_1$ zwischen zwei Netzwerkteilnehmern existiert genau dann, wenn Daten zwischen ihnen ausgetauscht werden können.

(b) Ein Link $L_1$ zwischen zwei Netzwerkteilnehmern existiert genau dann, wenn Daten zwischen ihnen ausgetauscht werden können und für alle Links $L_2$ mit $L_1 \neq L_2$ folgendes gilt.
$$\#L_1 \leq \#L_2$$

(c) Ein Link zwischen zwei Netzwerkteilnehmern existiert genau dann, wenn Daten zwischen ihnen ausgetauscht werden können und zwischen ihnen eine physikalische Verbindung existiert, welche diesen Datenaustausch im Rahmen des 802.11™ Standards ermöglicht.

Die Erweiterung der Linkdefinition nach (a) entspricht, wie oben beschrieben, dem Standard 802.11™ und ist sicherlich für viele Anwendungen hilfreich. Allerdings trifft dies nicht auf die Suche nach Brücken zu, da Brücken als Kanten zwischen zwei Knoten definiert sind, ohne welche ein Graph in mehrere Komponenten zerfällt. Das heißt, übertragen auf ein drahtloses Maschennetzwerk, Brücken sind die Verbindungen zwischen zwei Netzwerkteilnehmern, ohne welche das Netzwerk in mehrere Netzwerke zerfällt. Der Graph entspricht hierbei also dem Verbindungsgraphen und nicht etwa dem Erreichbarkeitsgraphen.

Dies ist auch die Begründung für das Verwerfen der Erweiterung (b). Selbst für eine kürzeste Liste von Netzwerkteilnehmern gilt, dass sie es ermöglicht, das zwei Netzwerkteilnehmer über einen dritten Netzwerkteilnehmer kommunizieren und derartige Links gilt es im Fall der Brückenerkennung nicht zu untersuchen.

Will man Brücken zwischen genau zwei Netzwerkteilnehmern erkennen, so sind Links als 1-Hop-Verbindungen zu betrachten, wie in (c) beschrieben. Eine physikalische Verbindung nach 802.11™ entspricht einer 1-Hop-Verbindung zwischen genau zwei Netzwerkteilnehmern, welche genau einer Kante in einem Verbindungsgraphen entspricht.

Nun ist es allerdings so, dass die Kommunikation zwischen zwei Netzwerkteilnehmern in einem drahtlosen Maschennetzwerk unzuverlässig ist. Das heißt, eine Kommunikation zwischen zwei Netzwerkteilnehmern kann spontan erfolgreich oder erfolglos sein. Nach der bisherigen Auffassung würde bei einer erfolgreichen Übertragung ein Link zwischen den zugehörigen Netzwerkteilnehmern existieren und andernfalls kein Link zwischen diesen Netzwerkteilnehmern existieren. Somit besteht der Bedarf an einem weiteren Kriterium für die hier zu untersuchenden Links, um Links trotz gelegentlicher erfolgloser Übertragungen bestimmen zu können.

Hierfür wird eine Metrik verwendet, um die Übertragungswahrscheinlichkeit zwischen zwei Netzwerkteilnehmern zu bewerten. Werden genügend viele Nachrichten erfolgreich übertragen, so verbessert sich die Bewertung der Übertragungswahrscheinlichkeit mit der verwendeten Metrik. Entspricht diese Bewertung einem zuvor festgelegtem Schwellwert, so existiert ein Link zwischen den beteiligten Netzwerkteilnehmern. Dieser Link bleibt auch erhalten, wenn gelegentlich Nachrichten nicht übertragen werden können, solange die Übertragungswahrscheinlichkeit insgesamt dem Schwellwert entspricht. Die Hierfür verwendete Metrik ist die ETX-Metrik [DC04].

**ETX-Metrik** Die ETX-Metrik, für **e**xpected **t**ransmission **c**ount, wurde 2004 im Rahmen einer Doktorarbeit am Massachusetts Institute of Technology entwickelt, um Routen in drahtlosen Maschennetzwerken zu bewerten. Bewertet wird eine Route anhand der Wahrscheinlichkeit $d_f$, dass eine Nachricht eines Absenders vom Empfänger empfangen wird und der Wahrscheinlichkeit $d_r$, dass eine Empfangsbestätigung des Empfängers vom Absender erhalten wird. Die Bewertung der Route zwischen dem Absender und dem Empfänger entspricht der folgenden Formel.

$$\frac{1}{d_f \times d_r}$$

Um die ETX-Metrik für die Bewertung von Verbindungen in einem drahtlosen Maschennetzwerk in unserem Anwendungsfall verwenden zu können, ist folgendes zu beachten. Die durch die ETX-Metrik zu bewertenden Routen haben eine Länge von eins, das heißt, jede Route besteht aus genau einem Hop. Diese 1-Hop-Routen entsprechen genau den 1-Hop-Verbindungen, womit die ETX-Metrik für die Bewertung von 1-Hop-Verbindung verwendet wird.

Weiter gilt zu beachten, dass während der DIBADAWN-Ausführung Unicast-Nachrichten und Broadcast-Nachrichten versendet werden. Möchte man mit Unicast-Nachrichten alle Verbindungen in einem drahtlosen Maschennetzwerk mit der ETX-Metrik vermessen, so muss jeder Knoten Unicast-Nachrichten an alle anderen Knoten im Netzwerk versenden. Da dies so nicht im DIBADAWN-Algorithmus vorgesehen ist, müsste sie zeitlich unabhängig vom DIBADAWN-Algorithmus durchgeführt werden und dies würde die Netzlast erhöhen, die DIBADAWN-Ausführung negativ beeinträchtigen und somit die Bewertung verfälschen.

Möchte man mit Broadcast-Nachrichten alle Verbindungen in einem drahtlosen Maschennetzwerk mit der ETX-Metrik vermessen, so muss jeder Knoten Broadcast-Nachrichten versenden. Da dies ohnehin durch die verteilte Ausführung des DIBADAWN-Algorithmus vorgesehen ist, würde keine zusätzliche Netzlast entstehen und die DIBADAWN-Ausführung würde somit nicht beeinträchtigt werden.

Eine Kombination beider Methoden ist auch möglich. Hierfür könnten alle Unicast- und Broadcast-Nachrichten für die Berechnung der ETX-Metrik verwendet werden. Allerdings erschwert dies die Auswertung. Nach dem Standard 802.11™ müssen Broadcast-Nachrichten, im Gegensatz zu Unicast-Nachrichten, nicht bestätigt werden. Dies führt dazu, dass sich die Berechnung der ETX-Metrik für Unicast- und Broadcast-Nachrichten unterscheidet. Durch das mögliche Verwenden der Unicast-Nachrichten würden die Verbindungen nur etwas intensiver vermessen werden, da Unicast-Nachrichten bei einer DIBADAWN-Ausführung lediglich in der Backward-Phase zwischen einem Netzwerkteilnehmer und einem weiteren Netzwerkteilnehmer, seinem parent im Suchbaum, versendet werden würden. Das bedeutet, es würden in einem drahtlosen Maschennetzwerk mit $n$ Netzwerkteilnehmern höchstens $n-1$ Verbindungen mit Unicast-Nachrichten vermessen werden. Im Gegensatz werden mit Broadcast-Nachrichten in der Forward-Phase alle benachbarten Netzwerkteilnehmer des Absenders erreicht, wodurch höchstens $n \times (n-1)$ Verbindungen vermessen werden.

Daher werden im Rahmen dieser Arbeit nur die Broadcast-Nachrichten betrachtet, um die ETX-Metrik zu berechnen. Die ausbleibende Bestätigungs-Nachricht beim Versenden von Broadcast-Nachrichten wird dabei so kompensiert, dass sich zwei potentiell verbundene Netzwerkteilnehmer für einen einheitlichen Zeitraum die Anzahl der selbst versendeten Broadcast-Nachrichten und die Anzahl der empfangenen Nachrichten pro Absender merken [DC04, S.57 f.]. Das heißt, wenn in einem Zeitraum ein Netzwerkteilnehmer A $tx_A$ Broadcast-Nachrichten versendet, wovon $rx_B$ von einem Netzwerkteilnehmer B empfangen werden. Und weiter B $tx_B$ Broadcast-Nachrichten versendet, wovon $rx_A$ von A empfangen werden. Dann entspricht die ETX-Metrik für die Verbindung zwischen A und B foldender Formel.

$$\frac{1}{\frac{rx_B}{tx_A} \times \frac{rx_A}{tx_B}}$$

Da die ETX-Metrik ausschließlich für die Bewertung des DIBADAWN-Algorithmus relevant ist, wird die ETX-Metrik für Verbindungen nicht zur Laufzeit eines Versuchs berechnet. Zur Laufzeit werden die gesendeten und empfangenen Broadcast-Nachrichten protokolliert, um die ETX-Metriken für alle Verbindungen nach der Beendigung eines Versuchs zu berechnen. Der für die Berechnung benötigte Zeitraum entspricht in diesem Fall dem gesamten Zeitraum der Versuchsausführung.

Abschließend ist im Kontext der Evaluation des Algorithmus DIBADAWN die Frage nach der Definition eines Links mit der Definition des ETX-Links beantwortet. Das heißt, anfänglich wird ein ETX-Schwellwert festgelegt und alle 1-Hop Verbindungen, welche diesem ETX-Schwellwert genügen werden als Link betrachtet.

## 6.2 Was ist die Referenztopologie?

Eine Referenztopologie eines Netzwerkes ist die tatsächlich in einem Netzwerk vorhandene Verbindungstopologie und beschreibt somit die Anordnung der Netzwerkteilnehmer und die Verbindungen zwischen diesen Netzwerkteilnehmern, welche als Referenz für die Bewertung dient. Bei der Evaluation des Algorithmus DIBADAWN liegt der Schwerpunkt auf den Verbindungen zwischen diesen Netzwerkteilnehmern. Daher wird in dieser Arbeit das Wort "Topologie" synonym zur Verbindungstopologie verwendet. Während bei einem kabelgebundenen Maschennetzwerk eine Topologie mit einer Verkabelung

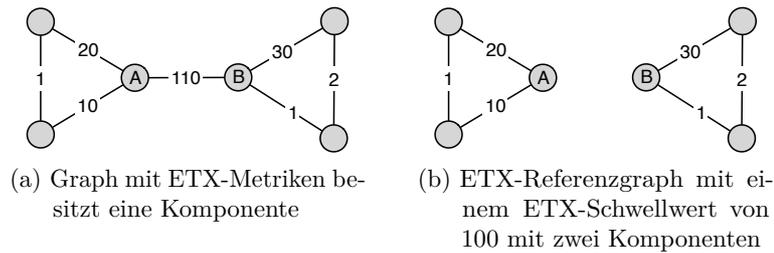

(a) Graph mit ETX-Metriken besitzt eine Komponente

(b) ETX-Referenzgraph mit einem ETX-Schwellwert von 100 mit zwei Komponenten

Abbildung 6.1: ETX-Referenzgraph aus einem Graphen

beziehungsweise mit einer Konfiguration naturgemäß vorliegt, ist in einem drahtlosen Maschennetzwerk eine Topologie nicht trivial zu bestimmen.

Für drahtlose Maschennetzwerke wird eine Verbindungstopologie definiert und dies kann auf mehrere Arten geschehen. Sie könnte beispielsweise auf der Basis des Datendurchsatzes, der Verbindungskosten oder anderen Metriken definiert werden. Im Rahmen dieser Arbeit wird die Verbindungstopologie und die Referenztopologie mit einem ETX-Referenzgraphen definiert. Für einen gegebenen Schwellwert $E_S$ existiert im drahtlosen Maschennetzwerk für alle ETX-Links mit ihren zugehörigen Netzwerkteilnehmern eine entsprechende Kante und ein entsprechender Knoten im ETX-Referenzgraphen. Dies bedeutet, der ETX-Referenzgraph besteht nicht zwingend aus einer einzelnen Komponente, wenn im drahtlosen Maschennetzwerk alle Netzwerkteilnehmer miteinander kommunizieren können. Die Abbildung 6.1 verdeutlicht dies beispielhaft. Im Graphen 6.1a befindet sich nur die Kante AB mit einer ETX-Metrik größer 100. Die Kommunikation über diese Kante ist demnach schwieriger, als über andere Kanten im Graphen. Für die Definition des ETX-Referenzgraphen in der Abbildung 6.1b wird ein ETX-Schwellwert 100 verwendet und daher wird die Kante AB nicht übernommen und der ETX-Referenzgraph beinhaltet mehrere Komponenten.

Bei der Nutzung eines ETX-Referenzgraphen als Referenzgraphen fällt auf, dass sich die erkannte Topologie für ein drahtlosen Maschennetzwerk ändern kann. Die ETX-Metrik-Werte einer Verbindung sind nicht statisch und können sich mit der Zeit ändern. Daher ist es für die Bewertung wichtig, die ETX-Metrik für die Verbindungen zeitnah zur DIBADAWN-Ausführung zu bestimmen. Wie im Abschnitt 6.1 beschrieben, werden die ETX-Metriken der Verbindungen auf der Grundlage gesendeter und empfangener Broadcast-Nachrichten berechnet. Dies bietet die Möglichkeit die tatsächlich zur Ausführungszeit vorherrschende ETX-Metrik zu erhalten. Diese Methode kann allerdings nur für eine nachträgliche Bewertung verwendet werden, da die Netzwerkteilnehmer in einem drahtlosen Maschennetzwerk, aufgrund des verteilten Ansatzes von DIBADAWN, nicht über die Sende- und Empfangszahlen der anderen Netzwerkteilnehmer informiert sind.

**Vorgehensweise**

Folgende Schritte werden durchgeführt, um einen ETX-Referenzgraphen mit einem Schwellwert $E_S$ zu erhalten.

1. Der DIBADAWN-Algorithmus wird verteilt wiederholt in einem Zeitraum ausgeführt.

2. Anhand der protokollierten Daten des zuvor ausgeführten DIBADAWN-Algorithmus werden die ETX-Metriken für die Verbindungen berechnet.

3. Der ETX-Referenzgraph mit dem Schwellwert $E_S$ wird berechnet.

4. Die Brücken und Gelenkpunkte im ETX-Referenzgraphen werden bestimmt.

## 6.3 Bewertung der Ergebnisse

In diesem Abschnitt wird die Bewertung der Ergebnisse beschrieben. Für die Bewertung werden die Brücken und Gelenkpunkte im ETX-Referenzgraphen bestimmt und mit den Brücken und Gelenkpunkten der einzelnen DIBADAWN-Ausführungen verglichen.

Bevor man eine geeignete Methode zur Bewertung der Ergebnisse wählt, ist es zielführend die Art der möglichen Ergebnisse zu betrachten. Bei der Ausführung von DIBADAWN werden Brücken und Gelenkpunkte gesucht und anschließend werden die gefunden Brücken und Gelenkpunkte mit der statistischen Komponente stabilisiert, wie im Algorithmus 2 in den Zeilen 2 und 3 zu sehen. Die Ausgabe ist eine Menge an Brücken und Gelenkpunkten.

In einem Graphen kann über jede Kante entschieden werden, ob diese Kante eine Brücke ist oder nicht. Dies kann ebenso für einen Gelenkpunkt entschieden werden. Über jeden Knoten im Graphen kann entschieden werden, ob er ein Gelenkpunkt ist oder nicht. Die Entscheidung, ob eine Brücke beziehungsweise ein Gelenkpunkt vorliegt, ist demnach in jeweils zwei Klassen geteilt. Eine Kante befindet sich entweder in der Klasse der Brücken oder der Klasse der Nicht-Brücken. Ebenso befindet sich ein Knoten entweder in der Klasse der Gelenkpunkte oder in der Klasse der Nicht-Gelenkpunkte. Der Algorithmus DIBADAWN liefert Ergebnisse in Form einer Menge an Brücken und Gelenkpunkten. Alle Kanten und Knoten, welche vom DIBADAWN-Algorithmus nicht als Brücken beziehungsweise als Gelenkpunkt erkannt werden, gelten als Nicht-Brücken beziehungsweise als Nicht-Gelenkpunkte.

**Klassifikation der Ergebnisse**

Bei der Bewertung des DIBADAWN-Algorithmus gilt zu beachten, dass die Ergebnisse der einzelnen DIBADAWN-Ausführungen mit einem ETX-Referenzgraphen verglichen werden. Wünschenswert ist eine möglichst gute Übereinstimmung des ETX-Referenzgraphen mit jeder Topologie, welche bei den einzelnen DIBADAWN-Ausführungen erkannt wird. Dies ist allerdings nicht zwingend der Fall. Werden nur die besten Links, also Verbindungen mit einer geringen ETX-Metrik, für die Definition

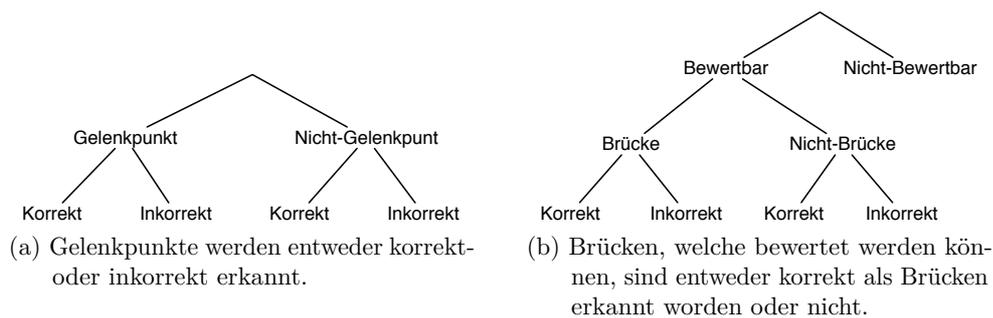

(a) Gelenkpunkte werden entweder korrekt- oder inkorrekt erkannt.

(b) Brücken, welche bewertet werden können, sind entweder korrekt als Brücken erkannt worden oder nicht.

Abbildung 6.2: Klassifikation bei der Bewertung

des ETX-Referenzgraphen verwendet, so werden möglicherweise viele Kanten nicht im ETX-Referenzgraphen erscheinen, welche vom DIBADAWN-Algorithmus ausgewertet wurden. Diese Kanten, unabhängig von ihrer Klassifizierung, können nicht bewertet werden, da der definierte ETX-Referenzgraph keine Aussage über diese Kanten enthält.

Somit wird entschieden, ob eine Bewertung einer Kante möglich ist. Ist dies der Fall, wird jede vom DIBADAWN-Algorithmus erkannte Brücke entweder in die Klasse der korrekt erkannten Brücken oder in die Klasse der inkorrekt erkannten Brücken eingeteilt. Die nicht vom DIBADAWN-Algorithmus benannten aber implizit gedeuteten Nicht-Brücken werden ebenfalls entweder in die Klasse der korrekt oder inkorrekt erkannten Nicht-Brücken eingeteilt, wie in der Abbildung 6.2b dargestellt.

Für Knoten ist dies praktisch ausgeschlossen, dass sie von einer DIBADAWN-Ausführung benannt werden, aber nicht im ETX-Referenzgraphen vorkommen, da der verteilte DIBADAWN-Algorithmus von allen Netzwerkteilnehmern im drahtlosen Maschennetzwerk gestartet wird und somit mindestens jeder Knoten Informationen über sich selbst liefern kann. Diese Informationen spiegeln sich ebenso im ETX-Referenzgraphen wieder, da dieser auch aus mehreren Komponenten bestehen kann. Daher ist die Anzahl der Elemente der Menge der Netzwerkteilnehmer gleich der Anzahl der Elemente der Menge an Knoten im ETX-Referenzgraph. Die Bewertung der Gelenkpunkte und implizit gedeutete Nicht-Gelenkpunkte ist ebenfalls in den Klassen "Korrekt" und "Inkorrekt" eingeteilt, wie in der Abbildung 6.2a dargestellt.

Die Unterteilung in bewertbare- und nicht bewertbare Brückenentscheidungen ist deshalb so wichtig, da diese Entscheidungen von DIBADAWN nicht bewertet werden und somit die Bewertung verfälschen können. Wenn verhältnismäßig viele DIBADAWN-Entscheidungen nicht bewertet werden, so hat offensichtlich die Bewertung nur einen geringen Wert, da viele DIBADAWN-Entscheidungen nicht berücksichtigt werden aber dennoch einen Einfluss im Netzwerk haben können.

Für alle Knoten und bewertbaren Kanten wird entschieden, ob sie Gelenkpunkte beziehungsweise Brücken sind und diese Entscheidung kann entweder als richtig oder falsch bewertet werden. Diese Art der Bewertung ist bekannt aus dem Information Retrival [Lew05] und lässt sich auch in einer Entscheidungstabelle darstellen, siehe Tablle 6.3.1. Die Entscheidungstabelle bewertet die Aussagen von DIBADAWN bezüglich folgender Fragen. Ist eine Kante eine Brücke? Ist ein Knoten ein Gelenkpunkt? Die Zeile "DIBADAWN" ist unterteilt in "ja" und "nein" für die jeweilige Antwort, welche der DIBADAWN-Algorithmus auf eine der beide Fragen liefert. Diese Antwort ist entweder richtig oder falsch, abhängig davon, ob der ETX-Referenzgraph diese Antwort unterstützt. In der Mitte ist die Anzahl der Entscheidungen aufgeführt in der jeweiligen Zeile und Spalte.

**Kennzahlen der Performanz**

Aus dieser Entscheidungstabelle lassen sich Kennzahlen bilden, welche die Performanz des DIBADAWN-Algorithmus unter verschiedenen Aspekten beschreiben. Sei TP, für "True Positive", die Anzahl der richtig positiven Entscheidungen und FP, für "False Positive", die Anzahl der falsch positiven Entscheidungen und FN, für "False Negative", die Anzahl der falsch negativen Entscheidungen.

$$
\begin{aligned}
\text{Precision} &:= \frac{TP}{TP+FP} \\
\text{Recall} &:= \frac{TP}{TP+FN} \\
F_1\text{-Measure} &:= 2\frac{\text{Precision} \times \text{Recall}}{\text{Precision} + \text{Recall}}
\end{aligned}
$$

Die Bedeutung der Precision und des Recalls wird exemplarisch für die Brücken beschrieben und gilt analog für die Gelenkpunkte. Die Precision gibt das Verhältnis zwischen korrekt erkannten Brücken, TP, und allen vom DIBADAWN-Algorithmus erkannten Brücken an. Wobei die korrekten Brücken immer anhand des ETX-Referenzgraphen bestimmt werden. Das Recall gibt das Verhältnis zwischen den korrekten, vom DIBADAWN-Algorithmus erkannten Brücken und allen korrekten Brücken an. Das $F_1$-Measure spiegelt das Verhältnis der Precision und des Recalls wieder, wobei die Precision und der Recall gleich stark gewichtet sind. Die Kennzahlen Precision, Recall und $F_1$-Measure beschreiben die Performanz des Algorithmus unter diesen Aspekten.

Allerdings sind je nach Anwendungszweck andere Aspekte denkbar. Beispielsweise könnte man sich vordergründig dafür interessieren, wie viele der, laut ETX-Referenzgraphen, tatsächlichen Nicht-Brücken vom Algorithmus erkannt werden im Verhältnis zu allen vom DIBADAWN-Algorithmus erkannten Nicht-Brücken. Um die Performanz zu beschreiben werden in dieser Arbeit hauptsächlich die Kennzahlen Precesion, Recall und $F_1$-Measure betrachtet. Daneben werden zusätzlich alle Werte der Entscheidungstabelle geliefert, um eine individuelle Bewertung nach den gewünschten Aspekten zu ermöglichen.

Die Schreibweise $F_1$-Measure wird üblicherweise verwendet, um auszudrücken, dass die Precision und das Recall zu gleichen Teilen in die Berechnung des F-Measures einfließen. Das dies in der vorliegenden Arbeit ausschließlich der Fall ist, wird synonym die Schreibweise F-Measure verwendet.

|  |  | ETX-Refenrenzgraph | |
|---|---|---|---|
|  |  | ja | nein |
| DIBADAWN | ja | richtig positiv (TP) | falsch positiv (FP) |
|  | nein | falsch negativ (FN) | richtig negativ (TN) |

Tabelle 6.3.1: Entscheidungstabelle

**Konfidenzintervall**

Die Kennzahlen werden berechnet, indem die Ergebnisse mehrerer DIBADAWN- Ausführungen über einen Zeitraum betrachtet werden. Um so mehr DIBADAWN-Ausführungen hierbei betrachtet werden, desto zuverlässiger sind die errechneten Kennzahlen. Dennoch ist es möglich, dass der gerade betrachtete Versuch die berechnete Performanz des DIBADAWN-Algorithmus überdurchschnittlich bevorteilt und somit das Ergebnis der Bewertung verfälscht.

Aus diesem Grund werden mehrere unterschiedliche Versuche betrachtet, wobei die berechneten Kennzahlen variieren können. Der Mittelwert dieser berechneten Kennzahlen stellt das Ergebnis der Bewertung dar. Dieser Mittelwert ist nicht als absolute Kennzahl zu betrachten, da dieser nur erhalten werden kann, wenn alle möglichen Versuche betrachtet werden, was aufgrund des unendlichen Eingaberaums nicht möglich ist. Daher wird zu dem Mittelwert noch das Konfidenzintervall angegeben. Das Konfidenzintervall beschreibt hierbei das Intervall, in welchem sich eine berechnete Kennzahl aus einem weiteren Versuch wahrscheinlich befinden wird.

Weiter ist für die Berechnung zu beachten, das das F-Measure nicht mit Mittelwerten berechnet werden darf. Da sich ein F-Measures, welches aus den Mittelwerten der Precisions und der Recalls berechnet wird, von einem Mittelwert aller F-Measure, welche direkt mit der Precision und dem Recall eines Versuchs berechnet werden, unterscheidet.

Um eine Kennzahl mit einem Mittelwert $X_M$ und einem Konfidenzintervall $[X_M - X_C, X_M + X_C]$ mit einer Wahrscheinlichkeit $X_P$, bei einer Versuchsanzahl $X_N$ zu beschreiben wird folgende Notation verwendet:

Die Kennzahl ist $X_M$ +/- $X_C$ ($X_P$% C.I., n=$X_N$).

# 7 Evaluation von DIBADAWN

Der hier evaluierte DIBADAWN-Algorithmus ist im Abschnitt 4 beschrieben. Die im Abschnitt 5 beschriebenen Probleme und Problemlösungen, welche bei der Implementierung aufgetreten sind, wurden ebenfalls in der evaluierten Variante berücksichtigt. Für die Evaluation werden die Versuche, wie im Abschnitt 5 beschrieben, im Simulator NS2 durchgeführt. Mit Hilfe der Simulation wurden bereits bei der Entwicklung mögliche Stärken und Schwächen des Algorithmus aufgedeckt, weshalb sich evaluierte optimale

Werte in den hier bereits in den nachfolgend beschriebenen Werten wiederfinden. Wie diese Werte ermittelt werden, ist ebenfalls in diesem Abschnitt beschrieben.

## 7.1 Generelles Vorgehen bei der Evaluation

In diesem Abschnitt wird das generelle Vorgehen bei einem Versuch festgehalten, so es nicht explizit anders beschrieben ist. Während eines Versuchs gibts es im Netzwerk keine Nutzlast außer der Nutzlast der DIBADAWN-Ausführung selbst. Hierbei gilt zu berücksichtigen, dass durch das Hinzukommen einer zusätzlichen Nutzlast die Kollisionswahrscheinlichkeit im Netzwerk wächst und daher die Übertragung von DIBADAWN-Nachrichten häufiger nicht erfolgreich ist und somit die Zuverlässigkeit des Algorithmus sinkt.

Im Rahmen dieser Arbeit werden zwei ETX-Schwellwerte verwendet. Zum Einen wird der ETX-Schwellwert 10 verwendet, dies entspricht einer symmetrischen Empfangswahrscheinlichkeit von zirka 31%. Das heißt, es werden für den ETX-Referenzgraphen Verbindungen verwendet, über welche Nachrichten mindestens mit einer Wahrscheinlichkeit von zirka 31% erfolgreich übertragen werden. Zum Anderen wird der ETX-Schwellwert 100 verwendet, dies entspricht einer symmetrischen Empfangswahrscheinlichkeit von 10%. Der ETX-Schwellwert 100 wird für alle Versuche verwendet, so kein anders ETX-Schwellwert explizit angegeben ist.

Für die Evaluation wird ein maximaler TTL-Wert für die zu übertragenden Nachrichten von 10 verwendet. Das heißt, DIBADAWN-Nachrichten werden über maximal 10 Hops übertragen. Wird eine DIBADAWN-Nachricht vom letzen Knoten empfangen, so entspricht der TTL-Wert der empfangenen Nachricht 0 und die Nachricht wird nicht weiter übertragen, wie im Algorithmus 2 in der Zeile 1 beschrieben.

Weiter werden 56 ms als MaxTraversalTime verwendet. Wie in den Abschnitten 4 und 5.3 beschrieben bedeutet dies, ein Netzwerkteilnehmer hat 56 ms Zeit, um eine DIBADAWN-Nachricht weiterzuleiten. Dementsprechend wird jitterMax auf 56/4 ms gesetzt, wie im Abschnitt 5.3 beschrieben.

Die Regeln Intelligente-Mehrheit und Kompetenter-Kreis treffen Entscheidungen auf der Basis von vertrauenswürdigen Aussagen, siehe Abschnitt 4. Bei Aussagen mit einer Kompetenzwahrscheinlichkeit von 100% ist davon auszugehen, dass diese vertrauenswürdig ist. Aber wie sieht es mit Werten nahe der 100% aus? Im Rahmen dieser Arbeit werden Aussagen mit einer Wahrscheinlichkeit ab 90% als vertrauenswürdig eingestuft. Weiter werden für das Voting jeweils die letzten 5 Aussagen betrachtet.

Weiter sei an dieser Stelle darauf hingewiesen, dass alle Dezimalzahlen in der englischen Schreibweise angegeben sind, also mit einem Punkt als Trennzeichen für Nachkommastellen.

**Zeitlicher Ablauf eines Versuchs**

Jeder Versuch ist so strukturiert, dass anfangs eine nicht evaluierte Phase durchlaufen wird und anschließend die Phase der Evaluation beginnt. Somit wird dem Algorithmus DIBADAWN die Möglichkeit gegeben, Informationen über das Netzwerk zu sammeln, welche von nachfolgenden DIBADAWN-Ausführungen, wie im Abschnitt 4 beschrieben, genutzt werden, um Entscheidungen zu treffen. Die Dauer der Aufwärmphase beträgt 200 Sekunden und die Dauer die evaluierten Phase beträgt 300 Sekunden. Alle 30 Sekunden wird von einem Knoten eine DIBADAWN-Ausführung mit einer gleich verteilten Wahrscheinlichkeit von 80% gestartet. Somit befinden sich die Resultate der letzen fünf DIBADAWN-Ausführungen im Speicher eines Netzwerkteilnehmers. Der Zeitpunkt der vermeintlich ersten Ausführung wird mit einer gleich verteilten Zufallszahl zwischen 0 und 20 Sekunden verzögert, um die Häufigkeit gleichzeitiger Starts mehrerer DIBADAWN-Ausführungen zu reduzieren. Die evaluierte Phase wird hierbei nicht abrupt beendet, sondern läuft mit der Beendigung der letzten Ausführung aus. Somit entspricht die Anzahl der evaluierten Ausführungen tatsächlich der Anzahl der vollständig ausgeführten DIBADAWN-Ausführungen. Die konkrete Dauer eines Versuchs und die Anzahl der Netzwerkteilnehmer kann variieren, jedoch sollen in jedem Versuch mindestens 2000 DIBADAWN-Ausführungen gestartet werden.

Wie im Abschnitt 6.2 beschrieben, wird während einer DIBADAWN-Ausführung ein Protokoll angefertigt, welches nach einem Versuch für die Evaluation ausgewertet wird. In diesem Protokoll wird unter anderem festgehalten, welche Brücken und Gelenkpunkte jedem Knoten momentan bekannt sind. Hierfür erstellt jeder Netzwerkteilnehmer alle 30 Sekunden einen Protokolleintrag. Jeder dieser Einträge eines Netzwerkteilnehmers ist mit einer laufenden Nummer versehen. Die Menge dieser Einträge im Protokoll mit der selben Nummer spiegelt den momentanen Wissensstand über Brücken und Gelenkpunkte im Netzwerk wider. Jeder dieser momentanen Sichtweisen sollte im Idealfall dem ETX-Referenzgraphen entsprechen, um eine möglichst gute Bewertung zu erzielen, wie im Abschnitt 6.3 beschrieben. Hierbei werden die Brücken und Gelenkpunkte der letzen DIBADAWN-Ausführung verwendet. Das heißt ein Voting findet im Regelfall nicht statt, so nicht anders beschrieben. Mit einem Voting können die Resultate noch verändert werden, wie beim Vergleich der Voting-Regeln gezeigt wird.

**Modulation And Coding Scheme**

Um Nachrichten in einem drahtlosen Maschennetzwerk von einem Netzwerkteilnehmer zu einem anderen zu übertragen, werden die Nachrichten in den unteren Schichten eines Netzwerks in ein physisches Signal umgewandelt. Laut dem WLAN-Standard 802.11™. gibt es verschieden Varianten eine Nachricht in solch ein Signal zu wandeln. Der Grund hierfür sind die unterschiedlichen Umgebungen, in welchen sich die Netzwerkteilnehmer befinden. Die Umgebung beeinflusst die Kommunikation erheblich. So können beispielsweise Störquellen Interferenzen erzeugen, welche sich negativ auf den Signal-Rausch-Abstand und somit auch negativ auf die Empfangswahrscheinlichkeit des Empfängers auswirkt. Weiter können sich beispielsweise Hindernisse auf die Dämpfung und somit ebenfalls negativ auf die Empfangswahrscheinlichkeit auswirken. Selbst der Abstand zwischen den Netzwerkteilnehmern fällt unter die Kategorie Umgebung und wirkt sich negativ auf die Dämpfung und somit auf die Empfangswahrscheinlichkeit des Empfängers aus. Um diesem Umstand Rechnung zu tragen, wird das zu übertragende Signal, das Nutzsignal, auf unterschiedliche Weise kodiert und modelliert.

Bei der Modulation wird die für die Übertragung genutzte physische Welle modelliert. Hierfür wird die Welle entweder mit einer Phasenmodulation oder mit einer Amplitudenmodulation oder beidem vom Absender modelliert. Abhängig von der Qualität des Übertragungskanals, also der Umgebung, können mehr oder weniger Abstufungen verwendet werden. Bei der Kodierung werden neben den eigentlichen Nutzdaten noch Daten für eine Vorwärtsfehlerkorrektur übertragen, sodass Bitfehler in den übertragenen Daten vom Empfänger korrigiert werden können und somit eine erneute Übertragung nicht notwendig ist.

Die möglichen Kombinationen aus einer Modellierungsart und einer Kodierungsart sind in den jeweiligen Teilspezifikationen des WLAN-Standards 802.11™ benannt und können dementsprechend verwendet werden. Für jede dieser spezifizierten Kombinationen lässt sich die maximal zu erreichende Datenrate berechnen, welche bei einer Übertragung von Nutzdaten unter Verwendung dieser Kombination erreicht werden kann. Im Rahmen dieser Arbeit und insbesondere für die Evaluation wird stets die feste Übertragungsrate von 1 MBit/s verwendet, so nicht explizit anders angegeben.

Auf die Nutzung eines Algorithmus zur dynamischen Wahl einer Übertragungsrate wurde verzichtet, da dies die Deutung der Evaluationsergebnisse erschwert hätte. Eine Veränderung der Evaluationsergebnisse aufgrund von unterschiedlichen Parametern könnte jeweils sowohl auf den DIBADAWN-Algorithmus als auch auf den anderen Algorithmus zurückzuführen sein.

**Wann gilt eine Brücke als erkannt?**

Im DIBADAWN-Algorithmus wird eine Brücke zweimal erkannt. Zu erst wird eine Brücke erkannt, indem ein Netzwerkteilnehmer keine Brücken- oder Kreis-Nachricht empfängt und somit über keine Nachrichten verfügt, welche er an seinen parent weitergeleitet kann. Dieser Netzwerkteilnehmer generiert daraufhin eine eigene Brücken-Nachricht und versendet diese an seinen parent. Der parent ist der zweite Netzwerkteilnehmer, welcher die Brücke erkennt, wie im Algorithmus 2 in der Zeile 4 zu sehen.

Bei der Übertragung dieser Brücken-Nachricht kann es, bedingt durch den unzuverlässigen Übertragungskanal, zu Übertragungsfehlern kommen. Für den Versand von Nachrichten an einen parent werden die Nachrichten als Unicast-Nachrichten versendet. Bei dieser Nachrichtenart wird vom Empfänger eine Übertragungsbestätigung innerhalb eines festen Zeitraums erwartet. Bleibt diese Bestätigungsnachricht aus, so wird die Übertragung der Nachricht wiederholt. Trotz der sieben Sendeversuche, beziehungsweise den sechs möglichen Wiederholungen, ist ein Nachrichtenverlust nicht auszuschließen.

Der Verlust dieser Nachricht führt dazu, dass der parent nicht über das Vorhandensein der Brücke informiert wird und sich aufgrund dessen anders verhält. Daher wird eine Kante nur genau dann als Brücke erkannt gewertet, wenn eine Brücke zwei mal innerhalb einer DIBADAWN-Ausführung erkannt wird.

Tabelle 7.2.1: Shadowing-Parameter

| Parameter | Wert | Parameter | Wert |
|---:|---|---:|---|
| $P_t$ | 24 | std_db | 2.0 |
| $G_t$ | 1.0 | dist0 | 1.0 |
| freq | $2.472 \times 10^9$ | RXThresh | -95.0 |
| L | 1.0 | CSThresh | -96.0 |
| pathlossExp | 2.0 | | |

## 7.2 Simulation

Eine Simulation hat den Vorteil, das viele verschieden Szenarien mit vergleichsweise wenig Aufwand konstruiert werden können. Die Anzahl, Einstellung und Position der Netzwerkteilnehmer eines drahtlosen Maschennetzwerks können beliebig konfiguriert und simuliert werden. Der Nachteil einer Simulation ist die bedingte Übertragbarkeit der Simulationsergebnisse auf die Wirklichkeit.

Möchte man etwas simulieren, so ist meist die Motivation einen komplexen Zusammenhang ergründen zu wollen. Hierfür wird zuerst durch die Beobachtung eines Phänomens ein Systemmodell erstellt, welches die elementaren zusammenhänge des Systems abstrakt beschreibt. Aus diesem Systemmodell wird ein Simulationsmodell konstruiert, um somit das Systemmodell formal zu beschreiben. Durch eine Ausführung des Systemmodells in einem Simulator werden Ausgaben erhalten, welche Rückschlüsse auf den komplexen zu betrachtenden Zusammenhang zulassen sollen. Die Schwierigkeit beim Modellieren und insbesondere beim Abstrahieren besteht nicht zuletzt darin, das System zu vereinfachen und dennoch alle relevanten Aspekte zu berücksichtigen. Ob die gewonnenen Ergebnisse auf die Wirklichkeit übertragbar sind, kann allerdings nicht generell beantwortet werden.

**Simulationsparameter**

Als Simulator wird der im Abschnitt 5 beschriebene Simulator NS2 mit dem Radio-Modell Shadowing verwendet. Ein Radio-Modell wird von NS2 verwendet, um die physische Signalstärke für jedes Paket beim Empfänger zu simulieren. Um mit Radio-Modell Shadowing eine gewünschten Umgebung zu modellieren, sind verschieden Einstellungen des Modells vorzunehmen. Die Parameter in der Tabelle 7.2.1 sind angegeben, um die Simulation und die Ergebnisse reproduzieren zu können.

Ohne die Simulations-Parameter konkret zu benennen, wären die Ergebnisse nicht mehr reproduzierbar. Dies soll am Beispiel des Parameter std_db verdeutlicht werden. Dieser Parameter beschreibt die Abweichung vom eigentlichem Pfadverlust in db und beeinflusst somit den Kurvenverlauf der Kurve für die Empfangswahrscheinlichkeit, wie in Abbildung 7.1 dargestellt. Es ist zu sehen, dass die Kurve bei einem Parameter std_sd = 0.5 zirka bei einer Entfernung von 190 Einheiten zu fallen beginnt bis die Kurve bei zirka 260 Einheiten bei ungefähr 0 ankommt. Damit fällt die Kurve der Empfangswahrscheinlichkeit für eine Standardabweichung von 0.5 am schnellsten. Schneller als die Kurve für die Standardabweichung von 2, welche für eine Entfernung im Bereich von 100 bis 400 Einheiten von 1 auf 0 fällt. Am schwächsten fällt die Kurve für eine Standardabweichung von 3.5. Sie fällt für eine Entfernung im Bereich von 50 bis über 500 Einheiten von 1 ab.

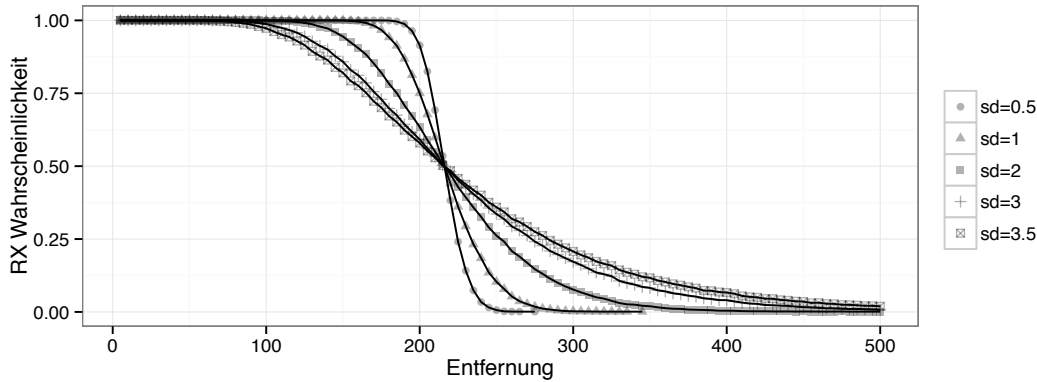

Abbildung 7.1: Empfangswahrscheinlichkeit bei variablen std_sd - Prameter

Die Darstellung der Kurve der Empfangswahrscheinlichkeiten ist für die Evaluation des DIBADAWN-Algorithmus deshalb so bedeutend, da sie mit den ETX-Links korreliert. Für die während der Evaluation verwendeten Standardabweichung von 2 ist die Empfangswahrscheinlichkeit im Bereich von 0 bis 100 Einheiten konstant 1. Das bedeutet, Knoten mit einer Distanz von 0 bis 100 Einheiten zu einem Absender empfangen wahrscheinlich alle Nachrichten dieses Absenders. Daraus berechnet sich eine ETX-Metrik von $1 = \frac{1}{1 \times 1}$, da die Verbindungen symmetrisch sind. Ab einer Entfernung von 100 Einheiten fällt die Empfangswahrscheinlichkeit ab und somit steigt die ETX-Metrik. Steigt der Wert für die ETX-Metrik einer Verbindung über den vorgegebenen Schwellwert, so wird diese Verbindung nicht für die Evaluation berücksichtigt, wie im Abschnitt 6 beschrieben. Wird beispielsweise ein ETC-Schwellwert von 100 angenommen für eine Standardabweichung von 2, dann werden nur Links mit einer Empfangswahrscheinlichkeit mit mindestens $0.1 = \sqrt{\frac{1}{100}}$ berücksichtigt, also mit einer Distanz von bis zu zirka 370 Einheiten. Demnach hätten alle Links mit einer Distanz im Bereich von 0 bis 100 Einheiten eine ETX-Metrik von 1, im Bereich von 100 bis 370 Einheiten eine ETX-Metrik zwischen 1 und 100. Würde man stattdessen eine Standardabweichung von 0.5 verwenden, so hätten alle Links mit einer Distanz im Bereich von 0 bis 190 Einheiten eine ETX-Metrik von 1, im Bereich von 190 bis zirka 250 Einheiten eine ETX-Metrik zwischen 1 und 100. Das heißt der Bereich für Metriken zwischen 1 und 100 wäre mit einer Standardabweichung von 2 mehr als vier mal so groß als mit einer Standardabweichung von 0.5. Somit kann bei der selben Anordnung der Netzwerkteilnehmer ein anderes Resultat erzielt werden, wenn die Parameter anders gewählt werden.

**Netzwerktopologie**

Möchte man Brücken und Gelenkpunkte in einem drahtlosen Maschennetzwerk finden, so benötigt man Netzwerke mit Brücken und Gelenkpunkten. Da für jeden Versuch eine Stichprobengröße von mindestens n=30 nach [Cra12, S. 123] angestrebt werden, werden pro Versuch mindestens 30 verschiedene Netzwerke für Simulationen benötigt.

Diese Netzwerke können nicht manuell erzeugt werden, da davon auszugehen ist, dass dies die Ergebnisse bewusst oder unbewusst verfälschen würde. Weiter würde dies die Wiederholbarkeit der Versuche verhindern.

Aus diesem Grund wird der NPART-Generator verwendet [MM09]. Hierfür werden Netzwerke mit 175 Netzwerkteilnehmern generiert, welche über eine maximale Empfangs- und Sendereichweite von 400 Einheiten verfügen. Weiter ist die *penalty* auf 15, *reduction* auf 0.2, *retries* auf 150 und der Topologietyp auf "Leipzig" gesetzt.

**TTL-Wert**

In einem drahtlosen Maschennetzwerk werden in der Forward-Phase DIBADAWN-Nachrichten ausgehend vom Initiator zwischen Netzwerkteilnehmern versendet, um einen Explorer im Netzwerk zu verbreiten, wie im Abschnitt 4 beschrieben. Hierbei werden allerdings nicht zwangsläufig alle Netzwerkteilnehmer erreicht. Gerade in großen Maschennetzwerken, würde dies zu viel Zeit in Anspruch nehmen und wäre aufgrund der unzuverlässigen Übertragung nur schwer zu erreichen. Wie im Algorithmus 2 in der Zeile 1 beschrieben, wird mit einer DIBADAWN-Nachricht ein Zähler übertragen, welcher nach dem Empfang vom Empfänger vermindert wird, wenn er nicht bereits den Wert 0 erreicht hat. Dieser Zähler nennt sich TTL für "Time to live". Der anfängliche TTL-Wert bestimmt somit die Anzahl der maximalen Übertragungen einer DIBADAWN-Nachricht entlang eines Übertragungsweges.

Die Verwendung solch einer Begrenzung ist zwar ein Bestandteil des Algorithmus, aber ein konkreter Wert für diesen Parameter ist nicht gegeben. Daher wird der anfänglich zu wählende TTL-Wert mit einer Simulation bestimmt. Hierfür wird die Auswirkung des zu wählenden TTL-Wertes auf die F-Measure ohne die Verwendung der Statistikkomponente betrachtet. Es wird ein Wertebereich von 2 bis 30 für TLL verwendet, da eine Brückenerkennung ohne eine Nachrichtenübertragung nicht funktionieren kann.

In der Abbildung 7.2 ist zu sehen, dass das F-Measure für die Brückenerkennung im TTL-Wertebereich 2 bis 10 von 0.424 +/-0.016 auf 0.565 +/-0.017 steigt, wobei für diese und alle anderen Werte dieses Versuchs ein Konfidenzintervall von 95% mit einer Stichprobengröße n=30 gilt. Ebenso befindet sich für die Gelenkpunkterkennung ein bei TTL=10 ein Hochpunkt mit einem F-Measure 0.402 +/-0.017, ob die Steigung zuvor von TTL=6 zu TTL=7 leicht fällt. Für den TTL-Wertebereich 11 bis 30 befindet sich für die Brückenerkennung bei TTL=15 ein Tiefpunkt mit dem F-Measure 0.551 +/- 0.021 und ein Hochpunkt bei TTL=29 mit dem F-Measure 0.566 +/-0.019. Für den TTL-Wertebereich 11 bis 30 befindet sich für die Gelenkpunkterkennung bei

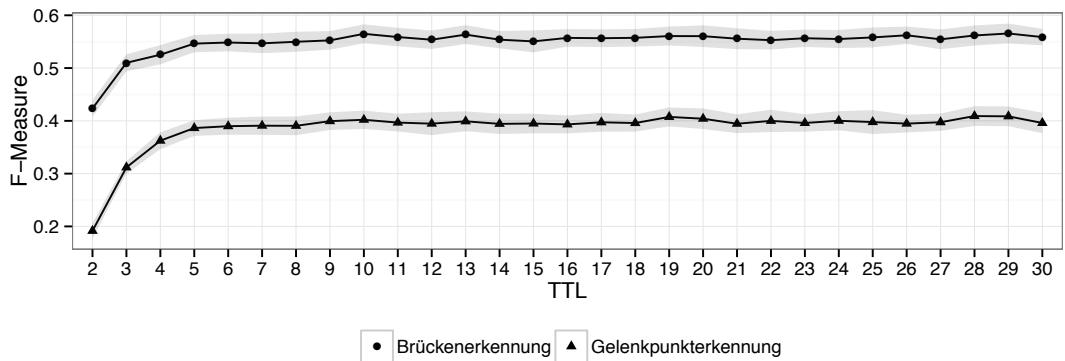

Abbildung 7.2: TTL und F-Measure

TTL=16 ein Tiefpunkt mit dem F-Measure 0.393 +/-0.017 und ein Hochpunkt bei TTL=28 mit dem F-Measure 0.409 +/-0.019. Für die Brückenerkennung im betrachteten TTL-Wertebereich von 2 bis 30 liegt das F-Measure bei TTL=29 nur 0.001 über dem F-Measure bei TTL=10. Für die Gelenkpunkterkennung im betrachteten TTL-Wertebereich von 2 bis 30 liegt das F-Measure bei TTL=28 nur 0.007 über dem F-Measure bei TTL=10. Dies bedeutet, um eine Verbesserung von 0.001 zu bewirken würde die Anzahl der maximalen Übertragungen einer DIBADAWN-Nachricht entlang eines Übertragungsweges beinahe verdreifacht werden.

Eine DIBADAWN-Nachricht wird zeitgesteuert weitergeleitet (siehe Abschnitte 4 und 5.3). Daher würde sich eine Verdreifachung des TTL-Wertes negativ auf die Laufzeit einer DIBADAWN-Ausführung auswirken. Sie würde sich ebenso verdreifachen. Weiter übersteigt die Varianz der angegebenen Werte die Differenz der Hoch- und Tiefpunkte, sodass die Verbesserung nicht bei allen Versuchen festzustellen ist. Aus diesem Grund wird im Rahmen dieser Arbeit der TTL-Wert 10 als Standard-Wert für alle Versuche verwendet, so nicht anders beschrieben.

Dieser Werte sind keines Falls allgemeingültig. Für jeden vermeintlichen optimalen TTL-Wert lassen sich Netzwerke konstruieren, welche die Leistung von DIBADAWN mit dem ermittelten Wert begünstigen und benachteiligen. Beispielsweise ein Netzwerk mit einer Ringtopologie, wobei jeder Knoten mit genau zwei anderen Knoten über eine sehr guten Link verbunden ist und das Netzwerk über genau $2 \times TTL + 1$ Verbindungen verfügt. In solch einem Netzwerk würde der DIBADAWN-Algorithmus nur falsche Brücken und Gelenkpunkte erkenne, da die Explorer-Nachricht in der Forward-Phase nie eine Querverbindung finden würde.

**MaxTraversalTime**

Während einer DIBADAWN-Ausführung wird in der Forward-Phase eine DIBADAWN-Nachricht von einem Knoten A mit einer h-Hop-Entfernung zum Initiator weitergeleitet, indem A ein Zeitfenster von MaxTraversalTime eingeräumt wird, um die Weiterleitung zufällig zu verzögern und die Nachricht per Broadcast zu übertragen, wie in den Abschnitten 4 und 5.3 beschrieben. Da eine DIBADAWN-Nachricht in der Forward-Phase maximal TTL-1 weitergeleitet wird und jedem weiterleitenden Netzwerkteilnehmer ein Zeitfenster von MaxTraversalTime für die Weiterleitung eingeräumt wird, beeinträchtigt der Wert von MaxTraversalTime die Ausführungsdauer einer DIBADAWN-Ausführung.

Das Weiterleiten innerhalb eines festen Zeitfensters ist zwar ein Bestandteil des Algorithmus, allerdings ist ein konkreter Wert für MaxTraversalTime nicht gegeben. Daher wird ein konkreter Wert für MaxTraversalTime mit einer Simulation ermittelt. Mit diesem Hintergrund wird die Auswirkung des Wertes MaxTraversalTime auf das F-Measure ohne die Verwendung der Statistikkomponente untersucht.

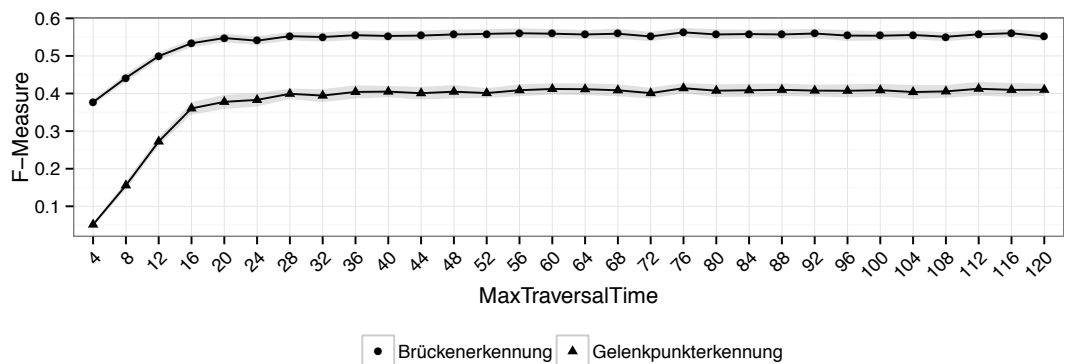

Abbildung 7.3: MaxTraversalTime und F-Measure

Wie in der Abbildung 7.3 zu sehen, steigt das F-Measure für die Brückenerkennung im MaxTraversalTime-Wertebereich von 4 bis 20 ms schnell von 0.377 +/-0.007 auf 0.548 +/-0.011, wobei für diese und alle anderen Werte dieses Versuchs ein Konfidenzintervall von 95% mit einer Stichprobengröße n=30 gilt. Ebenso steigt für die Gelenkpunkterkennung das F-Measure im MaxTraversalTime-Wertebereich von 4 bis 28 schnell von 0.051 +/-0.005 auf 0.399 +/-0.014. Im gesamten Wertebereich von 4 bis 120 ms wird ein Maximum für die Brückenerkennung von 0.563 +/-.0123 bei 76 ms erzielt. Ebenso wird ein Maximum für die Gelenkpunkterkennung von 0.414 +/-0.014 erzielt. Vor diesem Maximum ist bei der Brückenerkennung ein Hochpunkt bei MaxTraversalTime=56 ms erreicht mit einem F-Measure 0.560 +/-0.013 und liegt somit 0.003 unter dem genannten Maximum. MaxTraversalTime=56 erreicht die Gelenkpunkterkennung ein F-Measure von 0.409 +/-0.016 und liegt somit 0.005 unter dem beschriebenen Maximum.

Mit $(TTL-1) \times$ MaxTraversalTime wird die Ausführungsdauer der Forward-Phase berechnet, wobei TTL hierbei für den anfänglichen TTL-Wert steht. Dies bedeutet, dass sich eine Erhöhung der MaxTraversalTime mehrfach negativ auf die Ausführungsdauer einer DIBADAWN-Ausführung auswirkt. Daher werden im Rahmen dieser Arbeit 56 ms für MaxTraversalTime verwendet, um TTL × 10 ms pro DIBADAWN-Ausführung zu sparen.

**Linkdistanz**

Sind zwei Netzwerke über eine Brücke miteinander verbunden, so könnte man wollen, dass diese Brücke von einem angrenzenden Knoten erkannt wird, unabhängig davon welche weiteren Links dieser Knoten zu anderen Knoten in seinem angrenzenden Netzwerk besitzt.

Hierfür werden zwei Netzwerke, wie in der Abbildung 7.4 dargestellt, so platziert, dass genau ein Netzwerkteilnehmer des einen Netzwerkes mit genau einem Netzwerkteilnehmer eines anderen Netzwerkes über eine Brücke verbunden sind. Die an diese Brücke angrenzenden Netzwerkteilnehmer heißen A und B. Die Brücke AB hat eine Länge von 250 Einheiten. In jedem Netzwerk befinden sich genau 10 Netzwerkteilnehmer inklusive A beziehungsweise B. Weiter ist A und B so positioniert, dass sie eine Entfernung von genau $x_D$ zu allen anderen Netzwerkteilnehmern in ihrem Netz besitzen, dies entspricht den beiden Distanzen 1 und 2 in der Abbildung 7.4. Hierbei bildet die Distanz 1 mit 50 Einheiten den Abstand zwischen dem Brückenknoten und einer imaginären Geraden, welche eine Grenze für alle anderen Netzwerkteilnehmer im Netzwerk bildet, um weitere Verbindungen zwischen den beiden Netzen auszuschließen. Die Distanz 2 ist $x_D - 50$. Um zu verhindern, dass ein größerer Radius $x_D$ die Links zwischen den Netzwerkteilnehmern außer A und B zu sehr beeinflusst, ist die vertikale Ausdehnung auf dem Radius mit 300 Einheiten begrenzt. Dies ist mit der Distanz 3 in der Abbildung 7.4 dargestellt. Die Anordnung auf dem beschriebenen Radius ist gleich verteilt. Untersucht werden die Distanzen von 100 Einheiten bis 400 Einheiten. Dieser Bereich wurde gewählt, da nach der Abbildung 7.1 bei einer Distanz von 100 eine sehr gute Empfangswahrscheinlichkeit besteht und bei einer Distanz von 400 Einheiten eine sehr schlechte Empfangswahrscheinlichkeit vorliegt.

Da der Entfernung, wie oben beschrieben, eine Empfangswahrscheinlichkeit zugeordnet werden kann, könnte man davon auszugehen, dass eine Brücke zwischen Netzwerken mit einer kürzeren Linkdistanz zum Brückenknoten besser erkannt wird, da eine gestartete DIBADAWN-Ausführung mit einer höheren Wahrscheinlichkeit zum Brückenknoten erfolgreich übertragen wird, desto kürzer die Distanz zu diesem ist.

Die Links zwischen den anderen Netzwerkteilnehmern innerhalb eines der beiden Netze und die Links dieser Netzwerkteilnehmer zum entsprechenden Brückenknoten A oder B werden hier nicht untersucht und daher nach der Simulation und vor der Evaluation als Nicht-Bewertbar klassifiziert, wie im Abschnitt 6.3 beschrieben.

In der Abbildung 7.5 ist zu sehen, dass das F-Measure für die Brückenerkennung anfänglich 0.283 +/-0.015 bei einer Distanz von 100 Einheiten zwischen beiden Brückenknoten beträgt, wobei für diese und alle anderen Werte dieses Versuchs ein Konfidenzintervall von 95% mit einer Stichprobengröße n=31 gilt. Ab diesem Wert fällt das F-Measure bis zur Distanz von 160 Einheiten auf ein Tiefpunkt von 0.131 +/-0.031. Anschließend steigt das F-Measure bis zur Distanz von 320 Einheiten auf einen Wert von 0.518 +/-0.005 und übersteigt das anfängliche F-Measure bei einer Distanz von 100 deutlich. Für die Distanzen größer 320 Einheiten fällt das F-Measure anfangs leicht auf einen Tiefpunkt von 0.514 +/-0.006, erreicht bei 380 Einheiten das Maximum des Versuchs mit einem Wert von 0.523 +/-0.006 und fällt bei einer Distanz von 400 Einheiten wieder ab auf einen Wert von 0.513 +/-0.004.

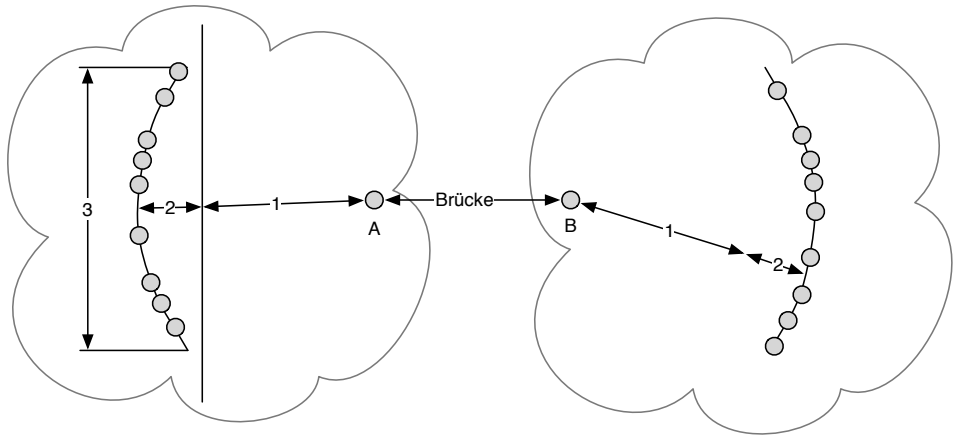

Abbildung 7.4: Versuchsaufbau

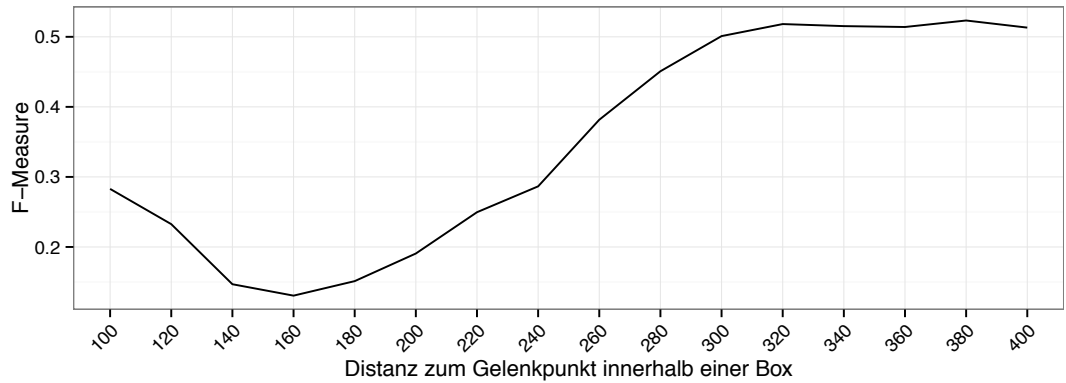

Abbildung 7.5: Linkdistanz und F-Measure

Somit wird die obige Überlegung bestätigt. Brücken mit angrenzenden Netzwerkteilnehmern mit weiteren guten Links erhöhen die Wahrscheinlichkeit der Brückenerkennung. Unerwartet hierbei ist allerdings, dass die Wahrscheinlichkeit der Brückenerkennung nur bis zu einer gewissen Distanz fällt und danach wieder steigt und dies sogar über den anfänglichen Wert mit den guten Links hinaus. In jedem Fall ist festzustellen, dass die Erkennung einer Brücke sehr wohl davon abhängig ist, welche weiteren Links die angrenzenden Knoten besitzen.

**Anzahl der Netzwerkteilnehmer**

Da nun bekannt ist, dass die Wahrscheinlichkeit der Brückenerkennung abhängig ist von der Linkqualität, welche die angrenzenden Knoten zu weiteren Knoten unterhalten, soll nun untersucht werden, ob ebenfalls eine Abhängigkeit von der Anzahl der zusätzlichen Links dieser Brückenknoten besteht. Da im vorherigen Versuch der Anstieg der Kurve ab einer Distanz von 300 Einheiten zurückging, könnte man vermuten, dass entsprechend der Abbildung 7.1 bei dieser Distanz weniger Nachrichten vom Brückenknoten empfangen wurden. Demnach könnte man erwarten, dass eine Abhängigkeit vom Knotengrad der angrenzenden Brückenknoten besteht.

Hierfür werden ebenfalls zwei Netzwerke, wie in der Abbildung 7.4 dargestellt, so platziert, dass genau ein Netzwerkteilnehmer des einen Netzwerkes mit genau einem Netzwerkteilnehmer eines anderen Netzwerkes über eine Brücke verbunden sind. Die an diese Brücke angrenzenden Netzwerkteilnehmer heißen A und B. Die Brücke AB hat eine Länge von 250 Einheiten. In jedem Netzwerk befindet sich mindestens ein Netzwerkteilnehmer inklusive A beziehungsweise B. Weiter ist A und B so positioniert, dass sie eine Entfernung von genau 230 Einheiten zu allen anderen Netzwerkteilnehmern in ihrem Netz besitzen. Dies entspricht den beiden Distanzen 1 und 2 in der Abbildung 7.4. Hierbei bildet die Distanz 1 mit 170 Einheiten den Abstand zwischen dem Brückenknoten und einer imaginären Geraden, welche eine Grenze für alle anderen Netzwerkteilnehmer im Netzwerk bildet, um weitere Verbindungen zwischen den beiden Netzen auszuschließen. Die Distanz 2 ist 60. Um zu verhindern, dass ein größerer Radius, also die Summe der Distanzen 1 und 2, die Links zwischen den Netzwerkteilnehmern außer A und B zu sehr beeinflusst, ist die vertikale Ausdehnung auf dem Radius mit 300 Einheiten begrenzt. Dies ist mit der Distanz 3 in der Abbildung 7.4 dargestellt. Die Anordnung auf dem beschriebenen Radius ist gleich verteilt. Untersucht werden unterschiedliche Anzahlen von Netzwerkteilnehmern in einem Netzwerk in Kombination mit einer unterschiedlichen Anzahl von Netzwerkteilnehmern im anderen Netzwerk. Die Anzahl der Netzwerkteilnehmer variiert zwischen 1 und 20.

Es werden ebenfalls die Links zwischen den anderen Netzwerkteilnehmern innerhalb eines der beiden Netze und die Links dieser Netzwerkteilnehmer zum entsprechenden Brückenknoten A oder B hier nicht untersucht und daher nach der Simulation und vor der Evaluation als Nicht-Bewertbar klassifiziert, wie im Abschnitt 6.3 beschrieben.

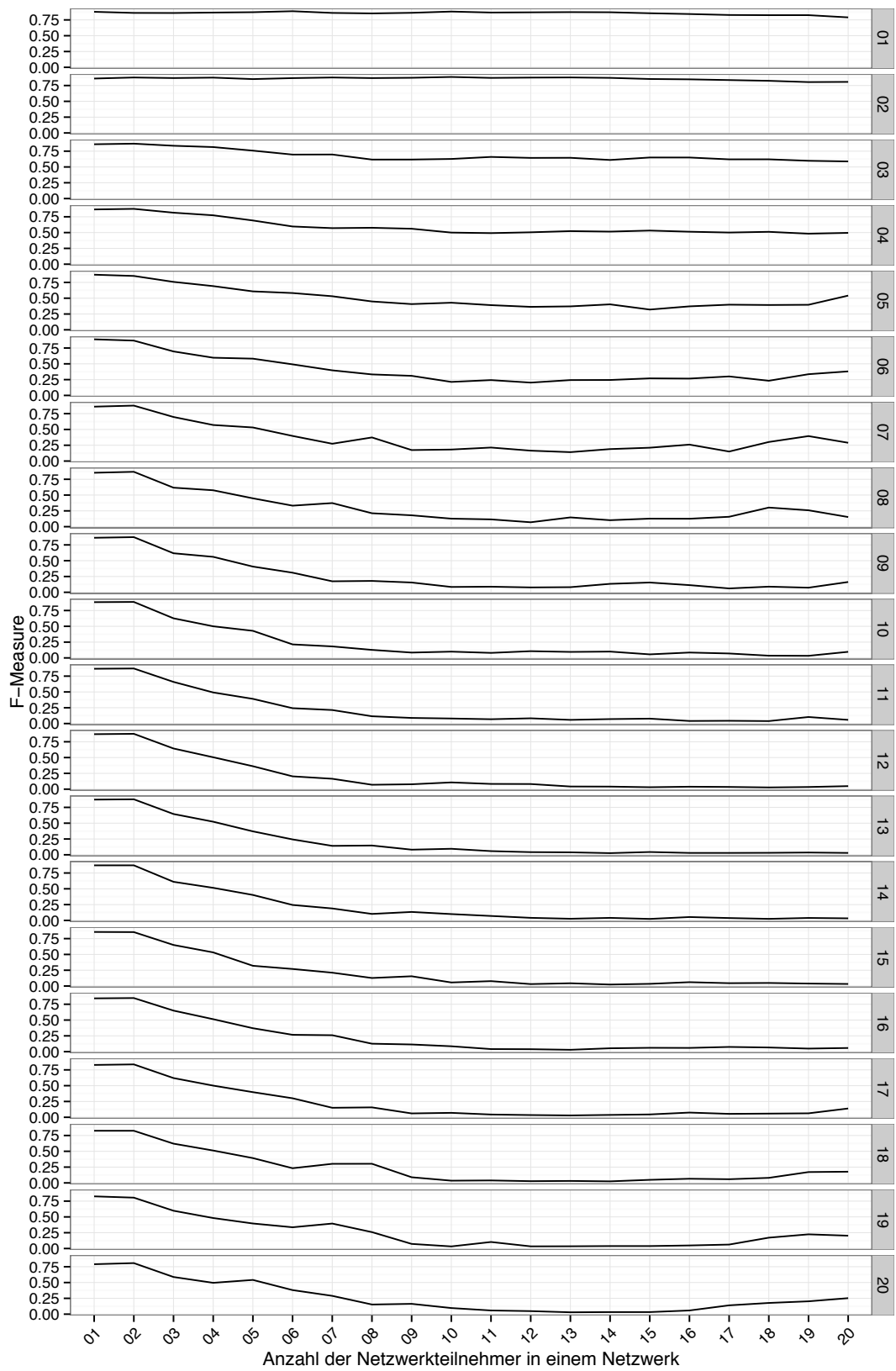

Abbildung 7.6: F-Measure bei der Kombination der Anzahl der Netzwerkteilnehmer

In der Abbildung 7.6 sind die Ergebnisse des Versuchs dargestellt. Die Anzahl der Netzwerkteilnehmer eines Netzwerkes befinden sich auf der rechten Seite und sind aufsteigend von oben nach unten sortiert. Diese Zahl bezeichnet eine Versuchsreihe mit einer konstanten Anzahl von Netzwerkteilnehmern in einem Netzwerk und einer variablen Anzahl von Netzwerkteilnehmern im anderen Netzwerk. Die Anzahl der Netzwerkteilnehmer des anderen, über eine Brücke verbundenen, Netzwerkes ist am unteren Rand abgetragen.

Befinden sich in beiden Netzwerken lediglich ein Netzwerkteilnehmer so wird ein F-Measure von 0.878 +/-0 erreicht, wobei für diese und alle anderen Werte dieses Versuchs ein Konfidenzintervall von 95% mit einer Stichprobengröße n=30 gilt. Wenn, sich in einem Netz 20 Netzwerkteilnehmer befinden, während im anderen weiterhin ein Netzwerkteilnehmer befindet reduziert sich das F-Measure auf 0.789 +/- 0.030. Diese Zahlen ähneln den Angaben der Versuchsreihe. Dies verhält sich anders für die Versuchsreihen größer zwei. In der Versuchsreihe 3 fällt das F-Measure von anfangs 0.859 +/-0.014 auf 0.587 +/-0.033 ab, wobei das F-Measure anfangs von einem Netzwerkteilnehmer zu zwei Netzwerkteilnehmern nicht so schnell fällt aber anschließend von 0.868 +/-0.023 auf 0.617 +/-0.049 und somit bereits beinahe das F-Measure von 20 Netzwerkteilnehmern erreicht. Die vierte Versuchsreihe ähnelt der dritten. Ab der fünften Versuchsreihe ist zu beobachten, dass das F-Measure zwischenzeitlich zum Schluss hin leicht verbessert, wobei diese Verbesserung nie an den Ausgangswert einer Versuchsreihe heranragt. Bei allen Versuchsreihen größer zwei fällt das F-Measure am stärksten im Bereich von zwei bis neun Netzwerkteilnehmern ab. Ab der Versuchsreihe 10 erreichen alle weiteren Versuchsreihen einen Tiefpunkt unter 0.05. Ab der 17. Versuchsreihe ist zu beobachten, dass F-Measure bei einer höheren Anzahl von Netzwerkteilnehmern wieder verbessert, so verbessert sich das F-Measure von 0.032 +/-0.024 bei einer Netzwerkteilnehmeranzahl von 14 konstant auf einen ein F-Measure 0.25 +/0.038 bei 20 Netzwerkteilnehmern.

Wie erwartet lässt sich eine Abhängigkeit vom Knotengrad der Brückenknoten beobachten. Diese Abhängigkeit ist nur schwach ausgeprägt, wenn der Knotengrad eines Brückenknoten gering ist und verstärkt sich, wenn der Knotengrad eines Brückenknoten steigt. Der drastische Abfall des F-Measures auf unter 0.05 ist hierbei besonders zu unterstreichen.

**Voting**

Während einer DIBADAWN-Ausführung trifft jeder an dieser Ausführung beteiligte Netzwerkteilnehmer Aussagen darüber, ob Verbindungen zu seinen Nachbarn Brücken sind und er selbst ein Gelenkpunkt ist. Aufgrund der Unzuverlässigkeit des Übertragungskanals und des damit verbunden Risikos Nachrichten erfolglos zu übertragen können diese Aussagen fehlerhaft sein. Um diese fehlerhaften Aussagen zu korrigieren zu, nutzt der DIBADAWN-Algorithmus die Statistikkomponente, wie im Abschnitt 4.6 beschrieben. Hierfür können unterschiedliche Regeln verwendet werden. Die Wirksamkeit dieser Regel soll mit diesem Versuch betrachtet werden indem ein Netzwerk mit jeweils einer unterschiedlichen Regel für die Brücken- und Gelenkpunkterkennung untersucht wird.

Für die gewichtete Regel wird, wie im Abschnitt 4.6 beschrieben, zusätzlich eine Wahrscheinlichkeit für das Vorkommen einer Brücke beziehungsweise eines Gelenkpunktes im Netzwerk benötigt. Diese Wahrscheinlichkeit wird vor dem eigentlichen Versuch ermittelt anhand der Anzahl der Brücken, Gelenkpunkte, Kanten und Netzwerkteilnehmer im entsprechenden ETX-Referenzgraphen. Dieser Graph wird hierbei durch eine vorherige Ausführung erhalten.

Die vom DIBADAWN-Algorithmus gefundenen Brücken und Gelenkpunkte werden mit den Brücken und Gelenkpunkten eines ETX-Referenzgraphen verglichen und die hieraus resultierenden Kennzahlen hier beschrieben. Als ETX-Schwellwert wird sowohl 10 als auch 100 verwendet, wie im Abschnitt 7.1 beschrieben. Weiter ist zu beachten die sie hier aufgeführten Werte Mittelwerte der Kennzahlen sind und diese nicht etwa Kennzahlen von Mittelwerten sind. Daher kann das hier dargestellte F-Measure nicht aus den hier dargestellten Precision und Recall berechnet werden.

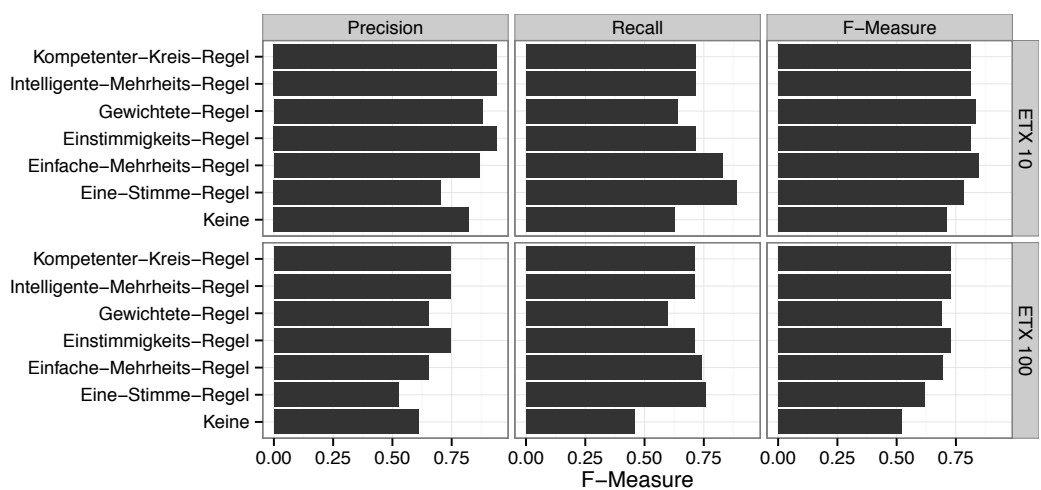

Abbildung 7.7: Vergleich der Voting Regeln unter der Verwendung unterschiedlicher ETX-Schwellwerte für die Brückenerkennung

**Brückenerkennung mit ETX 10** Wie in Abbildung 7.7 zu sehen wird sogar ohne die Verwendung der Statistikkomponente ein F-Measure von 0.710 +/-0.01 erzielt, wobei für diese und alle anderen Zahlen ein Konfidenzintervall von 95% und einer Stichprobengröße n=31 gilt. Diese F-Measure wird, wie im Abschnitt 6.3 beschrieben, berechnet mit der Precision und dem Recall. Die Precision 0.823 +/-0.02, welche zwar besser ist als die Precision der Eine-Stimme-Regel ist niedriger als die Precisions aller anderen Regeln. Das Recall 0.627 +/-0.016 ist niedriger als das Recall aller anderen Regeln. Im Vergleich zum F-Measure den anderen Regeln ist dieses F-Measure das geringste. Dies deckt sich mit der Motivation eine Statistikkomponente im DIBADAWN-Algorithmus zu verwenden.

Tabelle 7.2.2: Mittelwerte der Grundlegenden Kennzahlen für die Brückenerkennung mit einem ETX-Schwellwert 10

| TP | FP | TN | FN |
|---|---|---|---|
| 729 +/-27.3 | 45.3 +/-8.2 | 3872.16 +/-167.0 | 296.80 +/-24.6 |
| 849.3 +/-33.3 | 129.4 +/-17.3 | 3788 +/-165.9 | 176.41 +/-14.9 |
| 911.1 +/-36.6 | 384.5 +/-35.7 | 3532.96 +/-155.8 | 114.64 +/-8.2 |
| 729 +/-27.1 | 45.4 +/-8.1 | 3872.03 +/-167.1 | 296.80 +/-24.4 |
| 729.2 +/-27.3 | 45.4 +/-8.2 | 3872.06 +/-167.0 | 296.58 +/-24.6 |
| 645.6 +/-128.9 | 94.5 +/-37.6 | 3828.16 +/-169.8 | 377 +/-136.3 |
| 639.4 +/-26.2 | 138.8 +/-17.7 | 3783.86 +/-169.9 | 383.26 +/-27.3 |

Wie beschrieben ist die Precision der Eine-Stimme-Regel die Niedrigste mit 0.705 +/-0.024. Dies heißt 70.5% der vom DIBADAWN-Algorithmus gefundenen Brücken war eine Brücke im ETX-Referenzgraphen. Auf der einen Seite ist die Precesion mit Abstand die schlechteste aber auf der anderen Seite ist das Recall mit 0.889 +/-0.005 ebenso mit Abstand das beste Ergebnis. Dennoch erreicht die Eine-Stimmen-Regel nur ein F-Measure von 0.784 +/-0.015 und ist somit die schlechteste von den getesteten Regeln. Dies ist sicherlich nicht zuletzt in der hohen Varianz der TP- und FN-Kennzahl geschuldet, wie in der Tabelle 7.2.2 zu sehen. Das beste F-Measure 0.847 +/-0.009 wird mit der Einfache-Mehrheits-Regel erhalten, obgleich weder der Recall mit 0.829 +/-0.011 noch die Precision mit 0.868 +/-0.017 die höchste Kennzahl ausweist. Somit erzielt die Verwendung der Statistikkomponente mit der Einfache-Mehrheits-Regel eine Verbesserung des F-Measures von 0.137.

**Brückenerkennung mit ETX 100** Wie in Abbildung 7.7 zu sehen wird ohne die Verwendung der Statistikkomponente ein F-Measure von 0.521 +/-0.013 erzielt, was ebenfalls das geringste F-Measure darstellt. Wie mit dem ETX-Schwellwert 10, ist die Precision mit 0.611 +/-0.020 zwar höher als die Precision der Eine-Stimme-Regel aber geringer als die Precision aller anderen Regeln. Somit erscheint auch hier die Nutzung der Statistikkomponente gerechtfertigt.

Auffällig ist, dass sich der Recall aller Regeln deutlich verbessert. Selbst der Recall der gewichteten Regel hat sich auf 0.597 +/- 0.114 verbessert, wenngleich dies noch deutlich unter den Verbesserungen der restlichen Regeln bleibt. Die stärkste Verbesserung des Recalls erzielt die Eine-Stimme-Regel und erreicht den Wert 0.756 +/-0.010. Wie mit dem ETX-Schwellwert 10, fällt die Eine-Stimme-Regel auch hier dadurch auf, dass sie den höchsten Recall und die niedrigste Precision 0.527 +/-0.022 besitzt. Die Kompetenter-Kreis-Regel, die Intelligente-Mehrheits-Regel und die Einstimmigkeits-Regel besitzen sowohl eine ähnliche Precision von ungefähr 0.75, einen ähnlichen Recall von ungefähr 0.71 und sogar ein ähnliches F-Measure von ungefähr 0.72. Diese Ähnlichkeit ist auch für den ETX-Schwellwert 10 zu beobachten. Die gewichtete Regel liegt mit einem Recall 0.597 +/-0.114, einer Precision 0.651 +/-0.045 und einem F-Measure 0.688 +/-0.032 unter den eben beschrieben Regeln. Somit reichen die besten F-Measure Ergebnisse zwar nicht an die ETX-10-F-Measure Ergebnisse heran, stellen aber dennoch eine deutliche Verbesserung der Ergebnisse ohne die Statistikkomponente dar.

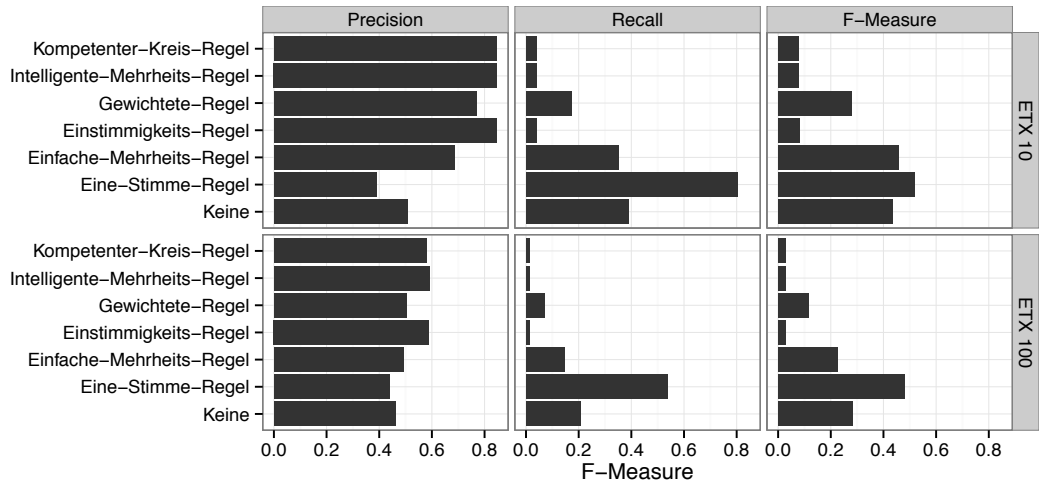

Abbildung 7.8: Vergleich der Voting Regeln unter der Verwendung unterschiedlicher ETX-Schwellwerte für die Gelenkpunkterkennung

**Gelenkpunkterkennung mit ETX 10** Wie in der Abbildung 7.8 zu sehen ist die Gelenkpunkterkennung weit weniger Erfolgreich als die Brückenerkennung. Mit dem ETX-Schwellwert 10 werden ohne Statistikkomponente 0.435 +/-0.019. Anders als bei der Brückenerkennung wird das F-Measure durch die Verwendung einer Regel nicht zwangsläufig verbessert. Die Eine-Stimme-Regel verbessert das F-Measure auf 0.519 +/-0.027 und die Einfache-Mehrheits-Regel verbessert das F-Measure geringfügig auf 0.459 +/-0.029. Hierbei ist, genau wie bei der Brückenerkennung, die Precision der Eine-Stimme-Regel die geringste mit 0.389 +/-0.028 und das Recall dieser Regel mit Abstand am größten mit 0.041 +/-0.008. Die Kompetenter-Kreis-Regel, die Intelligente-Mehrheits-Regel und die Einstimmigkeits-Regel besitzen eine ähnlich hohe Precision von gemittelt 0.694, einen ähnlich niedrigen Recall von gemittelt 0.041 und ebenfalls ein ähnliches niedrige F-Measure von gemittelt 0.079. Die Gewichtete Regel verfügt mit einer Precision 0.771 +/-0.040 und einem Recall 0.174 +/-0.028 über einem höheren F-Measure 0.277 +/-0.035 als die drei zuvor beschriebenen Regel, liegt aber dennoch deutlich unter dem F-Measure ohne Statistikkomponente.

Somit sind die dargestellten oberen vier Regeln nicht geeignet, um das F-Measure, also ein Verhältnis zwischen der Precision und dem Recall, zu steigern. Liegt das Augenmerk bei der Gelenkpunkterkennung jedoch auf der Precision, also auf der Richtigkeit der Erkannten Gelenkpunkte, und nicht etwa auf dem Recall, also der Vollständigkeit der Gelenkpunkterkennung, so sind die Kompetenter-Kreis-Regel, die Intelligente-Mehrheits-Regel und die Einstimmigkeits-Regel klar dafür geeignet diesen Aspekt zu verbessern. Ebenso lässt sich das Recall deutlich verbessern mit der Eine-Stimme-Regel.

**Gelenkpunkterkennung mit ETX 100**  Die Gelenkpunkterkennung mit einem ETX-Schwellwert von 100 vermindert das F-Measure um 35% auf einen Wert von 0.283 +/-0.019, wobei sich die Precision nur leicht auf 0.462 +/-0.031 verschlechtert, während sich das Recall drastisch auf 0.207 +/-0.018 verringert. Die Eine-Stimme-Regel kann auch hier das F-Measure verbessern auf einen Wert von 0.479 +/-0.019, was sogar über dem F-Measure ohne Statistikkomponente mit einem ETX-Schwellwert von 10 liegt, wobei das Recall auf 0.538 +/-0.025 verbessert. Die Kompetenter-Kreis-Regel, die Intelligente-Mehrheits-Regel und die Einstimmigkeits-Regel besitzen auch in diesem Fall eine ähnlich hohe Precision von gemittelt 0.586, einen ähnlich niedrigen Recall von gemittelt 0.015 und ebenfalls ein ähnliches niedrige F-Measure von gemittelt 0.029. Im Vergleich zu den drei eben genannten Regeln besitzen die Gewichtete-Regel und Einfache-Mehrheits-Regel eine geringere Precision, jedoch ein höheres Recall und ein deutlich höheres F-Measure, welches dennoch unter dem F-Measure ohne Statistikkomponente liegt.

Für die Verbesserung des F-Measures sind die oberen fünf Regeln nicht geeignet. Die Precision lässt sich leicht verbessern mit den Regeln Kompetenter-Kreis-Regel, Intelligente-Mehrheits-Regel und Einstimmigkeits-Regel, während das Recall mit der Eine-Stimme-Regel sogar mehr als verdoppelt werden kann.

Abschließend ist festzustellen, dass die Wirksamkeit der Statistikkomponente gegeben ist und je nach Fokus das Recall, die Precision oder andere Aspekte mit einer geeigneten Regel verbessert werden können.

# 8 Zusammenfassung

Es wurde gezeigt, dass sich die Wahrscheinlichkeit eine Brücke zu erkennen verbesserte, wenn angrenzenden Verbindungen kürzer als ein Schwellwert sind. Diese Verbesserung ist aber gering im Verhältnis zur Verbesserung, welche erzielt wird mit einer Verminderung dieser Verbindungen. Es wurde festgestellt, dass es einen Zusammenhang zwischen der Anzahl der Verbindungen eines an eine Brücke angrenzenden Netzwerkteilnehmers und der Wahrscheinlichkeit der Brückenerkennung gibt. Haben beide an eine Brücke angrenzende Netzwerkteilnehmer viele weitere Verbindungen, so wird die Wahrscheinlichkeit der Erkennung der Brücken geringer und umgekehrt steigt sie, wenn es weniger solcher Verbindungen gibt.

Weiter wurden Werte für die Anzahl der Maximalen Hops und für den Übertragungszeitraum für die Übertragung einer Nachricht zwischen zwei Netzwerkteilnehmern ermittelt. Diese Wert sind zwar keine Universalantwort für alle Netze, aber es ist beschrieben worden, wie diese Werte zu ermitteln sind und es ist zu vermuten, dass es eine obere Schranke für diese Parameter gibt.

Neben der eigentlichen Evaluation wurden Lücken der ursprünglichen DIBADAWN-Beschreibung aufgedeckt und bearbeitet. Dies ist insbesondere für zukünftige Implementationen von Bedeutung. So wurde in dieser Arbeit das verwendete Jitter genau beschreiben, die fehlerhafte Beschreibung der Kompetenzbewertung korrigiert, die Häufigkeit asymmetrischer Kreise gemindert und ein gleichmäßiges Verdrängen vergangener Aussagen beschrieben.

## 8.1 Ausblick

Der DIBADAWN-Algorithmus klassifiziert Verbindungen und Netzwerkteilnehmer als Brücken beziehungsweise als Gelenkpunkte. Eine Verbindung, exemplarisch betrachtet, ist entweder ein Brücke oder nicht, obgleich dieser Entscheidung innerhalb des DIBADAWN-Algorithmus eine Wahrscheinlichkeit für ihre Korrektheit zugeordnet ist. Es könnte untersucht werden, ob diese interne Information gewinnbringend nach außen gegeben werden kann. Das heißt, ob und in welchem Maß sich diese interne Wahrscheinlichkeit im vorliegenden Netzwerk widerspiegelt. Wenn sich diese interne Wahrscheinlichkeit nicht widerspiegeln sollte, so könnte weiter untersucht werden, wie man einen Algorithmus konzipieren müsste, um zusätzlich eine Wahrscheinlichkeit für jede getroffene Aussage zu erhalten.

Die Vorliegende Arbeit hat gezeigt, dass mit der Wahl guter Parameter die Leistungsfähigkeit von DIBADAWN verbessert werden kann. Somit ist die Wahl dieser Parameter entscheidend und es könnte untersucht werden, ob die Suche nach geeigneten Parametern vielleicht automatisiert werden könnte. Beispielsweise könnte eine Anwendung mit globalem Wissen über ein vorliegendes Netzwerk die DIBADAWN-Parameter regelmäßig nachjustieren. Möglicherweise könnte man die Leistungsfähigkeit von DIBADAWN verbessern, indem man die Abbildung für die Berechnung der Wahrscheinlichkeit der Korrektheit einer Kreis-Nachricht optimiert. Dies könnte ebenfalls mit dem vorherigen Vorschlag kombiniert werden.

Das Versenden von Broadcast-Nachrichten hat eine grundlegende Bedeutung für eine DIBADAWN-Ausführung, daher könnte untersucht werden, wie moderne Flooting-Algorithmen die Leistungsfähigkeit von DIBADAWN beeinflussen können.

## 8.2 Fazit

DIBADAWN ist ein Algorithmus für die Erkennung von Brücken und Gelenkpunkten in einem drahtlosen Maschennetzwerk. Es wurde festgestellt, dass die Brücken und Gelenkpunkte in Netzwerken mit guten- und weniger Verbindungen erkannt werden können, wobei der Erfolg dieser Erkennung an einer manuellen Justierung verschiedener Parameter wie MaxTraversalTime, MaxTTL und den zu nutzenden Regeln für die Stabilisierung der Ergebnisse gebunden ist.

# Literatur

**Selbständigkeitserklärung**

Ich erkläre hiermit, dass ich die vorliegende Arbeit selbständig verfasst und noch nicht für andere Prüfungen eingereicht habe. Sämtliche Quellen einschließlich Internetquellen, die unverändert oder abgewandelt wiedergegeben werden, insbesondere Quellen für Texte, Grafiken, Tabellen und Bilder, sind als solche kenntlich gemacht. Mir ist bekannt, dass bei Verstößen gegen diese Grundsätze ein Verfahren wegen Täuschungsversuchs bzw. Täuschung eingeleitet wird.

Berlin, den 23. Oktober 2014   ..............................................................................